\documentclass[preprint,12pt]{elsarticle}
\usepackage{setspace}

\usepackage{icarus}

\usepackage{natbib}

\bibpunct{(}{)}{;}{a}{,}{,}

\usepackage{graphicx}

\newcommand{\cotwo}{CO$_2$}
\newcommand{\htwoo}{H$_2$O}

\newcommand{\de}{$^{\circ}$}
\newcommand{\ms}{m~s$^{-1}$}
\usepackage{rotating}

\begin{document}
\doublespacing

\begin{frontmatter}



\title{An investigation of a super-Earth exoplanet with a greenhouse-gas atmosphere using a general circulation model}


\author[zalucha]{Angela M. Zalucha} 

\author[zalucha]{Timothy I. Michaels}
\author[madhu]{Nikku Madhusudhan}

\address[zalucha]{SETI Institute, 189 Bernardo Ave., Suite 100, Mountain View, CA, 94043, USA}
				
				\address[madhu]{Yale University, Department of Physics and Department of Astronomy, New Haven, Connecticut, 06511, USA}



%
%
%
%
%


\end{frontmatter}



\begin{flushleft}
\vspace{2.5cm} 
Number of pages: 80 \\
Number of tables: 1 \\
Number of figures: 17
\end{flushleft}


\begin{pagetwo}{GCM study of a greenhouse super-Earth}

\noindent Angela M. Zalucha\\
189 Bernardo Ave.\\
Suite 100\\
Mountain View, CA 94043, USA \\
\\
Email: azalucha@seti.org\\
Phone: (720) 208-7211

\end{pagetwo}

\begin{abstract}
We use the Massachusetts Institute of Technology general circulation model (GCM) dynamical core, in conjunction with a Newtonian relaxation scheme that relaxes to a gray, analytical solution of the radiative transfer equation, to simulate a tidally locked, synchronously orbiting super-Earth exoplanet.  This hypothetical exoplanet is simulated under the following main assumptions: (1) the size, mass, and orbital characteristics of GJ~1214b (Charbonneau et al., 2009), (2) a greenhouse-gas dominated atmosphere, (3), the gas properties of water vapor, and (4) a surface.  We have performed a parameter sweep over global mean surface pressure (0.1, 1, 10, and 100 bar) and global mean surface albedo (0.1, 0.4, and 0.7).  Given assumption (1) above, the period of rotation of this exoplanet is 1.58 Earth-days, which we classify as the rapidly rotating regime.  Our parameter sweep differs from Heng and Vogt (2011), who performed their study in the slowly rotating regime and using Held and Suarez (1994) thermal forcing.  This type of thermal forcing is a prescribed function, not related to any radiative transfer, used to benchmark Earth's atmosphere.  An equatorial, westerly, superrotating jet is a robust feature in our GCM results.  This equatorial jet is westerly at all longitudes.  At high latitudes, the flow is easterly.  The zonal winds do show a change with global mean surface pressure.  As global mean surface pressure increases, the speed of the equatorial jet decreases between 9 and 15 hours local time (substellar point is located at 12 hours local time).  The latitudinal extent of the equatorial jet increases on the nightside.  For the two greatest initial surface pressure cases, an increasingly westerly component of flow develops at middle to high latitudes between 11 and 18 hours local time.  On the nightside, the easterly flow in the midlatitudes also increases in speed as global mean surface pressure increases.  Furthermore, the zonal wind speed in the equatorial and midlatitude jets decreases with increasing surface albedo.  Also, the latitudinal width of the equatorial jet decreases as surface albedo increases.

\end{abstract}




\section{Introduction}\label{se:introduction}

Infrared observations of transiting  planets over the past decade have opened a window on a new regime of atmospheric dynamics~\citep[e.g.,][]{knutson:2007}.  The first transiting exoplanets discovered and modeled were ``hot Jupiters'', planets broadly similar in size to Jupiter that orbit their parent stars in close-in orbits (often with periods less than a few Earth-days) and hence receive an immense amount of stellar flux~\citep{mayor:1995,charbonneau:2000}.  Their masses and radii suggest that their atmospheres are hydrogen-dominated~\citep{fortney:2007}, similar to Solar System gas giants.  They are also expected to be tidally locked and synchronously orbiting their parent star, always keeping one face to their parent star.  These properties imply that their circulation patterns will be entirely unlike anything found in the Solar System.

\subsection{Review of exoplanet general circulation models}\label{ss:gcmsummary}

	Many exoplanet general circulation models (GCMs) have been developed, almost all from existing terrestrial GCMs.  \citet{showman:2002} developed the first true 3-D GCM of an exoplanet (in this case hot Jupiters such as HD~209458b) using the Explicit Planetary Isentropic Coordinate (EPIC) model~\citep{dowling:1998}.  This model differed from previous models of brown dwarfs in that they included the intense irradiation received from their parent star and differed from previous exoplanet models in that they included the effects of advection.  They predicted superrotating (westerly) winds, which later models have verified, and phase-curve offsets before the latter was discovered observationally.  \citet{cooper:2005} modeled HD~209458b using version 2 of the ARIES/GEOS dynamical core~\citep{suarez:1995} with Newtonian relaxation to a prescribed equilibrium temperature.  This model improved on \citet{showman:2002} by having a more realistic radiative-equilibrium temperature, better resolution, and a deeper domain.  It was thus able to better predict the day-night temperature difference.  \citet{cooper:2006} then used the \citet{cooper:2005} model to study the chemistry of HD~209458b.  CO and CH$_4$ were represented as passive tracers.  They found that where there is disequilibrium chemistry of CO and CH$_4$, their abundance also depends on the meteorology of the planet.

	Several hot Jupiter models have used the equivalent barotropic formulation or shallow water version of the primitive equations of atmospheric fluids.  \citet{cho:2003} developed a GCM for HD 209458b using the equivalent barotropic formulation of the shallow-water equations with Newtonian heating to a prescribed equilibrium state and obtained different results than \citet{showman:2002}.  Namely, they predicted a banded structure with three broad zonal jets and easterly flow at the equator.  \citet{menou:2003}, in a companion paper, generalized the \citet{cho:2003} GCM to other giant planets by varying the Rossby and Burger numbers.  They found that close-in giant exoplanets lie in a different regime of flow patterns than Solar System gas giant planets.  Namely, the former have few zonal jets, while the latter have many.  \citet{cho:2008} extended the study of \citet{cho:2003} to a much larger parameter space by varying the initial RMS velocity, global mean temperature at cloud tops, thermal forcing amplitude, Rossby deformation radius at the pole, and nondimensional Rhines length at the equator.  They concluded that giant exoplanets have a polar vortex that revolves around each pole, a low (two or three) number of zonal jets (in contrast with Solar System gas giants), and a temperature field that depends on the circulation.  \citet{rauscher:2008} used the GCM of \citet{cho:2008} to predict observational signatures of hot Jupiters.  \citet{langton:2007} used a shallow water model similar to \citet{cho:2003} and found a cold spot centered 76\de~east of the antistellar point.  They attributed this disagreement to a difference in their radiative time constant.

		Meanwhile, hot Jupiter models with other types of dynamical cores continued to develop and improve.  \citet{showman:2008} simulated the atmospheres of the hot Jupiters HD 209458b and HD 189733b using a GCM based on the Massachusetts Institute of Technology (MIT) GCM dynamical core, which uses the full 3-D privative equations.  The \citet{showman:2008} GCM was radiatively forced using Newtonian relaxation to a radiative equilibrium state.  \citet{showman:2009} continued this work using a radiative transfer (RT) scheme~\citep{marley:1999} that calculated the RT fluxes directly using a correlated-k scheme with multiple wavelengths in the two-stream approximation.  Both of these studies showed flow from the substellar to antistellar point along the equator and over the poles at lower pressures, while a westerly equatorial jet and easterly polar flow developed at higher pressures.  \citet{menou:2009} used the Intermediate GCM dynamical core of \citet{hoskins:1975}, which solves the primitive equations.  This GCM was thermally forced by Newtonian relaxation to a prescribed equilibrium temperature and used to compare the barotropic (hot Jupiter) and baroclinic (Earth-like) regimes.  \citet{rauscher:2010} used the same model as \citet{menou:2009} to check that hot Jupiters were being modeled consistently with the results of \citet{cooper:2005,cooper:2006} and \citet{showman:2008,showman:2009}.  \citet{perna:2010} used the \citet{rauscher:2010} GCM to show that magnetic drag is a plausible mechanism to limit wind speeds in hot Jupiters; \citet{perna:2010b} used the \citet{rauscher:2010} GCM to show that Ohmic dissipation is a non-negligible heat source in hot Jupiters.  The \citet{menou:2009} GCM was subsequently  improved by \citet{rauscher:2012} to include dual band, double-gray RT.
		
		Several studies have investigated the effects of particular aspects of hot Jupiter GCMs on the resulting circulation and temperature.  \citet{burkert:2005} used an inviscid, primitive equation model in 2-D (varying in azimuth and radius) with flux limited radiative diffusion to show that the nightside temperature is sensitive to atmospheric opacity.  Similarly, \citet{dobbs-dixon:2008} used a flux limited radiative inviscid hydrodynamical model (truncated at $\pm$70\de~latitude) to show that atmospheres with large opacity have a large day-night temperature difference, while atmospheres with reduced opacity have a more uniform temperature.  \citet{dobbs-dixon:2010} improved on the model of \citet{dobbs-dixon:2008} by using decoupled radiative and thermal components with updated opacities.  They also added viscosity, and found that high viscosity resulted in subsonic flow, while low viscosity produced supersonic flow.  The addition of viscosity changed the circulation patterns and variability of flows.  \citet{langton:2008} used a shallow water model with two a two-wavelength radiative transfer to investigate the effect of high orbital eccentricity on hot Jupiters.  They found that the atmospheric response was driven primarily by the intense irradiation at periastron and that the resulting expansion of heated air produced high velocity turbulent flow (with some planets developing superrotating acoustic fronts as well).  Finally, \citet{thrastarson:2010}, using a GCM with the primitive equations and Newtonian relaxation to a prescribed equilibrium state, found that different initial states of the initial zonal jet (which had the most significant effect), thermal drag timescale, spectral truncation number, and superviscosity coefficient, lead to markedly different results for circulation and temperature.  \citet{thrastarson:2011}, using the same GCM, found that a short Newtonian relaxation time led to large amounts of unphysical, gird-scale oscillations that contaminated the flow field.
		
		Heng, Menou, and Phillipps (2011) performed benchmark simulations of Earth, tidally locked synchronously orbiting Earth, a shallow hot Jupiter, and a deep HD~209458b.   The purpose of benchmark simulations is to compare dynamical cores of GCMs to determine if differences are due to numerical modeling procedures or ``higher level'' physical parameterizations.  Heng, Menou, and Phillipps (2011) followed the standard Earth method of \citet{held:1994} with a slight modification to include the effects of a tidally locked synchronously orbiting planet.  This scheme used Newtonian relaxation to relax to the following radiative equilibrium temperature:
		\begin{equation}\label{eq:hs}
		T_{eq,hs}=max\left\{T_{strat},\left[T_{surf} - \Delta_h \cos\left(\lambda-180^{\circ}\right)\cos\phi-\Delta_z log\left(\frac{p}{p_o}\right)cos^2\phi\right]\left(\frac{p}{p_s}\right)^{R/c_p}\right\},
		\end{equation}
		where $T_{eq,hs}$ is the \citet{held:1994} equilibrium temperature (in K), $T_{strat}$=200~K is the stratospheric temperature, $T_{surf}$=315~K is the surface temperature, $\Delta_h=60$~K is the equator to pole temperature difference, $\lambda$ is longitude, $\phi$ is latitude, $\Delta_z=10$~K is a characteristic temperature difference in altitude, $p$ is pressure, $p_s$ is the surface pressure (here taken to be $10^5$~Pa), $R$ is the specific gas constant, and $c_p$ is the specific heat at constant pressure.  Since \citet{held:1994} were comparing terrestrial dynamical cores, they set their parameters in Eq.~\ref{eq:hs} to mimic the temperature distribution in Earth's atmosphere.   Heng, Menou, and Phillipps (2011), in their benchmark test, compared the Geophysical Fluid Dynamics Laboratory (GFDL) - Princeton flexible modeling system (FMS) (see references in Heng, Menou, and Phillipps 2011) spectral and finite difference cores.  They found qualitative and quantitative agreement between the two cores in all but the deep HD~209458b case, where horizontal dissipation changed the results with errors of $10\%$ or greater in the temperature and wind fields.  Heng, Frierson, and Phillipps (2011) continued the development of the GFDL-Princeton FMS exoplanet model by adding dual-band radiative transfer, dry convective adjustment, and a uniform slab of finite thickness to model the surface.  Their goal was to bridge the purely dynamical 3-D simulations with those that incorporate multiband radiative transfer.  In addition, \citet{perna:2012} used the Heng, Menou, and Phillipps (2011) and Heng, Frierson, and Phillipps (2011) models to show that on hot Jupiters, irradiation by the parent star is the main driver in radiative and dynamical properties, but that atmospheric opacity also plays an important role.
		
		Thus far in our discussion, we have focused on Jupiter-size planets or larger, but models for smaller planets exist too.  The main commonality for all of these models is that the planet is tidally locked with and synchronously orbiting its parent star.  There seems to be a continuum in size for these planets, at least in the modeling sense, ranging from Jupiter size or larger, to Neptune size, to mini-Neptunes, to super-Earths (two to ten Earth masses), to Earth or smaller.  Moreover, there is a also a range of modeling choices on how to specify the bottom boundary condition (in terms of the heat and momentum exchange), which will be noted throughout Section~\ref{se:gcm}.

Turning our attention to smaller-sized planets, \citet{lewis:2010} used the \citet{showman:2009} GCM to study the hot Neptune GJ~436b and found that as atmospheric metallicity was increased, the circulation changed from one dominated by mid-latitude jets and minimal longitudinal temperature differences, to one dominated by an equatorial jet and large day-night temperature differences.  \citet{menou:2012} used the \citet{rauscher:2012} GCM to model super-Earth GJ 1214b with \htwoo, solar, and $\times$30 super solar metallicity and found that superrotating equatorial jets were robust features for these cases.

Long before the discovery of super-Earth exoplanets, \citet{joshi:1997} used a terrestrial planet GCM (i.e. shallow atmosphere covering a thick rocky surface), based on the simple GCM of \citet{joshi:1995} and \citet{james:1986} to simulate a tidally locked, synchronously rotating exoplanet.  They used a two stream, gray RT representing a gas with \cotwo~or \htwoo~radiative properties.  The model was somewhat of a blend of terrestrial (radius, gravitational acceleration at the surface, gas constant, and stellar insolation) and Mars constants (for instance, the atmosphere could freeze and sublimate according to \cotwo~parameters), but also had a slow, Titan-like rotation rate (16 Earth-days).  Their results contained a longitudinal, thermally direct cell, which transported heat from the dayside to the
nightside. The mass was returned to the dayside through the polar regions at low levels.  In the \citet{joshi:1997} model, the circulation aloft was fairly longitudinally symmetric, and a superrotating jet was
present at the equator.  The winds were westerly almost everywhere.

\citet{joshi:2003} followed up the work of \citet{joshi:1997} by switching to the Intermediate GCM described in \citet{forster:2000} and adding more model features such as non-gray radiative transfer; clouds; continents and oceans; and evaporation, precipitation, and transport of water.  They found that the presence of a dry continent caused higher temperatures on the dayside and allowed accumulation of snow on the nightside.  Furthermore, the absence of any ocean led to higher day-night temperature differences.

\citet{merlis:2010} used an idealized, terrestrial aquaplanet, primitive equation, spectral GCM~\citep{ogorman:2008} that contained an active hydrologic cycle.  They used a two stream RT scheme that had a shortwave component and two gray long wave components (one where optical depth depended linearly on pressure to represent a well mixed gas such as \cotwo, and the other with a quartic dependence representing \htwoo).  They performed two simulations, assuming the planet was tidally locked to and synchronously orbiting its parent star: one in which the rotation period was equal to one current
Earth year and the other in which the rotation period was equal to one current Earth day.  They found that waves and eddies shaped the winds, temperature, and precipitation in the rapidly rotating atmosphere, while simple divergent circulations dominated the slowly rotating atmosphere.

Heng and Vogt (2011) used the Heng, Menou and Phillipps (2011) GCM to simulate super-Earth Gliese~581g.  They varied the radiative cooling time, the rotational frequency of the planet, the mass of the planet, the density of the planet, the equator to pole temperature difference, the global mean surface pressure, and presence or lack of a troposphere.  They found the radiative cooling time to have the most substantial effect on the results.  A long radiative time scale compared with the advection time scale allowed the circulation to diffuse the temperature field.  \citet{selsis:2011} used the LMD GCM \citep[see references in ][]{selsis:2011} to simulate a tidally locked, sychronously orbiting super-Earth sized planet with a \cotwo~atmosphere to determine the wavelength and phase changes in the thermal emission as atmospheric pressure increased.

A few studies have investigated the stability of liquid water on super-Earths.  \citet{pierrehumbert:2011} modeled Gliese~581g with the FOAM GCM \citep[previously used to study snowball Earth and other deep-time paleoclimate problems; see references in][]{pierrehumbert:2011}.  They found that their ``Eyeball Earth'' simulations, i.e. mostly frozen planets with a pool of liquid water centered at the substellar point could maintain liquid \htwoo~stability at the surface.   \citet{wordsworth:2011} simulated a different super-Earth, Gleise~581d, using the LMD Mars model \citep[see references in][]{wordsworth:2011} to determine if a \cotwo~atmosphere was stable against atmospheric collapse and if liquid water could be stable on the surface.  They found this is the case for surface pressures with over $10^6$~Pa of \cotwo~for both land and ocean surfaces and varying amounts of background gas.

\subsection{Present Study}

The motivation for our work came with the transit observation of super-Earth GJ~1214b by \citet{charbonneau:2009}.  \citet{zalucha:2010} and \citet{zalucha:2012} had previously adapted the MIT atmosphere GCM to Pluto and Mars, respectively.  Given the success of the Mars and Pluto versions, which we call collectively the MIT small body GCM, it seemed reasonable to extend these GCMs to the super-Earth regime with solid or liquid surfaces.  Note that this is a separate effort from \citet{showman:2009} who use the giant planet version that is derived from the MIT ocean GCM.

  Recent spectral observations of GJ~1214b~\citep{bean:2010,bean:2011,desert:2011,crossfield:2011,berta:2012} indicate that the atmospheric composition is either metal rich or hydrogen-dominated~\citep{croll:2011,demooij:2012}.  \htwoo~and \cotwo~have been suggested as possible candidates for the metal rich case.  However, \citet{rogers:2010} have shown using planet interior structure models, that if GJ~1214b is composed of water, it must include a super-critical fluid layer.  Such a layer is not currently present in the MIT small body GCM.  A separate study by \citet{nettelmann:2011} favored a Neptune-type planet with an envelope (i.e. atmosphere) composed of $\sim$85\% water by mass mixed into H and He atop a rock core
with about 0.2$\times$ solar bulk ice to rock ratio.

Given the vast amount of exoplanets and exoplanet candidates that are being discovered at present, we chose to simulate a hypothetical planet that has (1) the size, mass, and orbital characteristics of GJ~1214b; (2) a greenhouse-type atmosphere (i.e., an atmosphere whose heating/cooling properties are dominated by a greenhouse gas without scatterers, akin to a dust-free Mars); (3) the gas properties of \htwoo~(but without any sort of surface ocean or hydrologic cycle); and (4) a surface, i.e. some density discontinuity between a gaseous atmosphere and solid or liquid surface that provides a source of friction and interacts thermally with the atmosphere.  We investigate a range of global mean surface pressures ($10^4$--$10^7$~Pa) and surface albedos (0.1--0.7), since these are not yet constrained by observations.  

As shown by our literature review above, exoplanet GCMs are abundant.  We use a state of the art dynamical core compared with \citet{joshi:1997}.  As previously mentioned, \citet{showman:2009} use the same dynamical core that we do, but use different parameterizations and more importantly lack a surface.  \citet{menou:2012} has modeled GJ~1214b with a \htwoo~atmosphere, but in the mini-Neptune regime since it does not have a solid or liquid surface.  This is simply a modeling choice; we compare the differences between our model and theirs in Section~\ref{ss:comparison}.  \citet{merlis:2010} perform idealized simulations, but only for rapidly vs. slowly rotating cases (their rapidly rotating case will be useful to compare to our model).

Heng and Vogt (2011) have already performed parameter sweeps of super-Earth planets, but they use Newtonian relaxation to the \citet{held:1994} radiative equilibrium state (Eq.~\ref{eq:hs}).  \citet{held:1994} has been and continues to be a widely-used, robust, workhorse for terrestrial GCMs.  Although it has tunable parameters for other planets (e.g., equator to pole temperature difference), its \textit{functional form} is deliberately set so that it implicitly mimics the features of Earth's atmosphere.  We use the same technique of Newtonian relaxation, but our expression for the radiative equilibrium temperature is an actual solution to the RT equations (and includes convection).  Granted, the solution is not completely state-of-the-art in that it is a gray model, but simple models are not without value (more complex does not always mean more accurate).  Furthermore, no parameter sweep like ours has been performed using any of the models with sophisticated RT.

Another important difference from the Heng and Vogt (2011) study is that their default case has a rotation period of 37 Earth-days, while ours assumes the GJ~1214b rate of 1.58~Earth-days.  \citet{merlis:2010} have noted the importance in rotation rate, to the extent that the circulation is in entirely different regimes for 1 Earth-day periods and 365 Earth-day periods.  We speculate in Section~\ref{ss:comparison} that the rotation period of 16 Earth-days in the \citet{joshi:1997} GCM is also in the slow regime.  Heng and Vogt (2011) themselves even state that rotation rate is one of the most important parameters that affect the atmospheric circulation.  Thus, as Heng and Vogt (2011) performed a parameter sweep in the slowly rotating regime, we perform a parameter sweep in the rapidly rotating regime.
	
In Section~\ref{se:gcm} we describe the 3-D GCM dynamical core (based on the MIT GCM) and external heating/cooling scheme of Newtonian relaxation to a prescribed radiative-convective equilibrium temperature profile.  In Section~\ref{se:results} we present the results for temperature, winds, mass streamfunction, and eddy transport, focusing on the $10^5$~Pa global mean surface pressure and 0.4 albedo case. We also perform a comparison of temperature, winds, and surface pressure variation as global mean surface pressure and albedo are varied.  In Section~\ref{ss:sp11}, we compare our results to the shallow water model of \citet{showman:2009} and other 3-D GCMs in Section~\ref{ss:comparison}.  We also present model transmission spectra in Section~\ref{ss:spectra}.

	\section{General circulation model configuration}\label{se:gcm}
	\subsection{Dynamical core}
	
		\citet{zalucha:2010} adapted the MIT GCM to Mars and used it to analyze the effect of topography on the Martian Hadley cells.  \citet{zalucha:2012} also adapted the MIT GCM to Pluto and compared the results to observed stellar occultation light curves.  Other groups have also adapted the MIT GCM dynamical core to hot Jupiters and hot Neptunes~\citep{showman:2009,lewis:2010}.  Thus the MIT GCM has been successfully applied to other planets beyond its already wide usage as an Earth atmosphere and ocean model.
		
		The dynamical core of the MIT GCM solves the primitive equations using a finite volume method on an Arakawa C-grid~\citep{marshall:1997} on a sphere.  The model atmosphere is hydrostatic (see the Appendix for justification of this assumption), compressible, and behaves according to the ideal gas law.  A thermodynamic equation is included that contains a user-specified external heating term (see Section~\ref{ss:rs}).  Convective adjustment is performed within the atmosphere to prevent superadiabatic temperature profiles.    We assume that the gravitational acceleration is constant with height.   In the horizontal, a cubed-sphere grid~\citep{adcroft:2004} with 32 $\times$ 32 points per cube face is used, equivalent to a grid spacing of 2.8\de~or 834~km at the equator.  
		
		The Rossby radius of deformation in the midlatitudes is given by 
		  \begin{equation}
		  L_{Ro}=\frac{Nh}{f_o},
		  \end{equation} 
		  where $N\sim 0.01$~s$^{-1}$ is the Brunt-–V\"{a}is\"{a}l\"{a} frequency, $h\sim 10$~km is the order-of-magnitude scale height, and $f_o \sim 10^{-4}$~s$^{-1}$ is the midlatitude Coriolis parameter.  Then, $L_{Ro}$ is of order 1000~km.  The Rhines length is given by
		\begin{equation}
		 L_{Rhines}=\left(\frac{U}{\beta}\right)^{1/2},
		 \end{equation}
		  where is $U$ is a characteristic velocity (i.e., the square root of the globally averaged total kinetic energy), taken to be 100~\ms, and $\beta$ is the meridional derivative of the Coriolis parameter, taken to be $5\times10^{-12}$~(m~s)$^{-1}$.  The Rhines length is then approximately 4000~km.  Thus, the horizontal model grid resolves both the midlatitude Rossy radius of deformation and Rhines length.

		Compared to a cylindrical projection grid (i.e., a latitude/longitude grid), this type of horizontal grid eliminates singularities at the poles that force meridional winds to zero and removes the requirement for Fourier filtering in the high latitudes (in order to maintain a practical timestep).  Note that in contrast to the finite volume method employed by the MIT GCM, other exoplanet GCMs (e.g. Rauscher and Menou 2012; Heng and Vogt 2011; Cho et al., 2003) use the spectral method, which allows for a cleaner control of numerical noise.  The MIT GCM configuration has neither viscosity nor vertical diffusion, but uses a Shapiro S2 filter to remove purely numerical noise~\citep{shapiro:1970}.  An 8th order filter is used for the momentum.  The order of the physical space filter is 4.  The timescale of the filter is set to model time step.  Heng, Menou, and Phillipps (2011) have demonstrated that numerical horizontal dissipation renders the prediction of temperatures to be uncertain at the 10\% level and several tens of percent in wind speeds in hot Jupiter models; however this was not the case in their Earth models.  Comparison of our derived wind speeds to that of other studies may suffer from this uncertainty.
		
		The vertical grid is defined in terms of an $\eta$ coordinate~\citep{adcroft:2004} based on atmospheric pressure and is scaled according to the prescribed global mean surface pressure.   $\eta$ coordinates have the advantage of being based in atmospheric pressure (which reduce the governing equations to a simpler form than height coordinates) and do not suffer large pressure gradient errors near the surface when the topography is non-zero (as with a terrain-following or $\sigma$ coordinate).  In general, surfaces of constant $\eta$ will intersect the surface, making computation more difficult, though for our purposes we assume flat topography and this is not a concern.   The vertical grid has 30 points that vary logarithmically in pressure, such that there are more points closer to the surface.  Throughout most of the model domain, there are about 5--10 points per scale height, except near the bottom where the resolution is finer and the top where the resolution is coarser.  The top level of the model is located at $\eta=2\times10^{-5}$; however, the top three points of the model contain Rayleigh drag (discussed below) and are not considered physically realistic output.  The top most level of the model with physically realistic output is located at $\eta=10^{-3}$.

			The horizontal momentum equation is given by:
\begin{equation}
\frac{D\mathbf{u}_h}{Dt} = -f\mathbf{\hat{k}}\times\mathbf{u_h}-\nabla_h\Phi+\mathbf{F}_h,
\end{equation}
where $D$ is the material derivative, $t$ is time, \textbf{u}$_h$ is the horizontal component of velocity, $f=2\Omega\sin \phi$ is the Coriolis parameter (where $\Omega$ is the planetary rotation rate), $\Phi$ is the geopotential height, $\nabla_h$ indicates that only the horizontal component of the gradient should be considered, and \textbf{F}$_h$ is the frictional drag.  Here friction arises from drag on the atmosphere by the surface, given by \citep[following][]{held:1994}
\begin{equation}\label{eq:bls}
\mathbf{F}_h  =  -k_f\max\left(0,\frac{p-p_b}{p_o-p_b}\right)\mathbf{u}_h\label{eq:bldrag}
\end{equation}
where $k_f$ is the wind damping rate,  $p_o$ is the local surface pressure (a function of latitude, longitude, and time), and $p_b=0.7p_o$ is the pressure at the top altitude of the boundary layer, defined as the part of the atmosphere where the drag is nonzero.  Eq.~\ref{eq:bldrag} represents a drag law (linearly dependent on the horizontal velocity) that decreases with height, reaches zero at the top of the boundary layer, and is zero at all levels above.  It reflects the diminishing dynamical influence of the surface the altitude increases and the fact that the boundary layer tends to follow the height of the surface terrain (i.e., the top of the boundary layer will generally be higher where the surface elevation is higher).  \citet{held:1994}, which is a benchmark GCM for Earth, uses $k_f=1$~Earth-day$^{-1}$.  Mars models use frictional time scales of 0.2 to 10 sols~\citep{joshi:1995}, 1~sol~\citep{lewis:1996}, 0.25-0.7~sol~\citep{nayvelt:1997}, and $\sim$20~sols~\citep{caballero:2008}.  These relationships are determined empirically through modeling of near-surface atmospheric properties.  Earth and Mars are the only bodies for which empirical values exist, and in the lack of any data for exoplanets, we take them to be representative of terrestrial-like exoplanets and use $k_f=1$~day$^{-1}$.

Note that Eq.~\ref{eq:bls} does not represent a true boundary layer scheme, but mimics one using Rayleigh friction.  See~\citet{garratt:1994} for a review of boundary layer science for Earth.  Furthermore note that the method of dealing with the momentum at the bottom boundary of hot Jupiter models has also not yet reached a consensus.  Heng, Frierson, and Phillipps (2011) use a free-slip condition, i.e., no friction at the bottom boundary and no condition that the velocity parallel to the surface should be zero.  On the other hand, \cite{rauscher:2012} use Rayleigh friction at the bottom boundary to mimic the effects of magnetic drag.

Rayleigh friction is also used at the model top as a parameterization for small-scale waves that break (i.e. deposit momentum) in the upper atmosphere~\citep{leovy:1964}.  Rayleigh friction is a drag force linearly dependent on velocity, similar to the surface friction scheme above, but the force increases with height in the uppermost layers of the model.  Rayleigh friction schemes are tuned to atmospheric measurements where observations of density, temperature, and/or wind exists (i.e. Earth, Mars, and Venus), but for exoplanets no information regarding small-scale waves is available.  Thus, in the top three levels we use typical Earth drag coefficients, from top to bottom these are 1, 3, and 9 Earth-day$^{-1}$.

Table~\ref{tb:gcmall} shows the GCM parameters.  We have assumed that the gaseous properties are that of a 100$\%$ water vapor atmosphere.  This choice is somewhat arbitrary, as we could have chosen another greenhouse-type gas such as CO$_2$.  Nonetheless, we do not expect the fundamental conclusions to be drastically different.  Although we are not explicitly modeling GJ~1214b, we have chosen to use the observed parameters of this planet from \citet{charbonneau:2009}.  Along with the assumption of a solid surface, this combination of parameters represents a hypothetical planet, since \citet{rogers:2010} showed that GJ 1214b cannot have a solid surface if its atmosphere is composed of water vapor.

	      \citet{heller:2011} have shown that planets in close orbits around low-mass stars will lose their obliquities on very short timescales.  Provided that putative other planets in the system do not disturb
the planet significantly, it will not have a significant obliquity, even it was initially
close to 90\de~(R. Heller, private communication).  Thus, we assume zero obliquity, which by definition means the planet will be at perpetual equinox and have no seasons.  \citet{heng:2012} have shown that the characteristic time scales for spin synchronization and circularization of the orbit of GJ 1214b (whose orbital parameters we use) is probably much less than the age of the star.  Thus, we assume a circular orbit.

	\subsection{Newtonian relaxation scheme}\label{ss:rs}

 In our GCM the planet is assumed to be tidally locked into a synchronous orbit with its parent star.  Like the Mars version of our GCM, we use the method of Newtonian relaxation to prescribe the external heating/cooling term in the model, given by
\begin{equation}\label{eq:newtonian}
c_p\frac{DT}{Dt}-\frac{1}{V}\frac{Dp}{Dt}=-k_T (T-T_{eq,c})
\end{equation}
where $T$ is the air temperature in the GCM, $c_p$ is the specific heat at constant pressure, $V$ is volume, $T_{eq,c}$ is the radiative-convective equilibrium air temperature, and $k_T$ is the radiative relaxation rate.  $k_T$ is a parameter that is calculated based on the $e$-folding time of a perturbation to the vertical temperature profile~\citep[the method is described in detail in Eq. 6 of][]{showman:2008}.  We note that several studies (Heng and Vogt, 2011; Showman and Polvani, 2011; Langton and Laughlin, 2007; Thrastarson and Cho, 2011) have pointed out the importance of radiative relaxation rate on the atmospheric circulation, but assume constant $k_T$ here for simplicity.  

In deriving the expression for $T_{eq,c}$, we first start with the radiative equilibrium temperature $T_{eq}$ (and its value at the surface $T_{eq,s}$) that does not contain the effects of convection; we refer to it as the \textit{pure} radiative equilibrium temperature.  For an atmosphere that is gray at long wavelengths and transparent at short wavelengths, the RT equation in the Eddington approximation has the simultaneous analytical solution for the atmosphere
\begin{equation}\label{eq:teq}
\sigma T^4_{eq}=\left(1-A\right)Q\left(0.5+0.75\tau\right)
\end{equation}
and surface
\begin{equation}\label{eq:teqs}
\sigma T^4_{eq,s}=\left(1-A\right)Q\left(1+0.75\tau_s\right),
\end{equation}
where $\sigma$ is the Stefan-Boltzmann constant, $A$ is the surface albedo, $\tau$ is the optical depth, $\tau_s$ is the optical depth at the surface, and $Q$ is the stellar flux received by the planet at a particular latitude $phi$, hour angle $h$ (related to longitude), and stellar declination $\delta$ (related to season).  $Q$ is given by,
\begin{equation}
Q=Q_o\cos\left(\sin\phi\sin\delta + \cos\phi\cos\delta\cos h\right),
\end{equation}
where $Q_o$ is the stellar flux received by the planet (equivalent to the solar constant defined for Earth).  This radiative equilibrium model predicts the surface temperature $T_{eq,s}$ by flux balance with the outgoing surface flux ($\sigma T^4_{eq,s}$), the downwelling long wave flux from the atmosphere, and the downwelling short wave flux from the star.  As such, the only assumption about the properties of the surface is that it have a density discontinuity with the atmosphere (e.g., is a liquid or solid).  The surface is infinitesimally thin and has zero heat storage capacity (i.e. no thermal lag) in this radiative equilibrium model (although we could account for an emissivity not equal to unity by multiplying the left hand side of Eq. \ref{eq:teqs} by the emissivity).  Note that the surface temperature of the GCM output may be prognostically determined using a slab method or other method of the user's choosing.

The optical depth is linearly dependent on pressure as  $\tau= \tau_{oo}(p/p_{oo})$, where $\tau_oo$ \citep[whose numerical value is taken from][]{merlis:2010} is the optical depth at the reference pressure $p_{oo}=10^5$~Pa.  Thus, as the surface pressure is varied, so will $\tau$ at the surface, indicating a stronger greenhouse effect for more massive atmospheres.  Our gray atmosphere specification is equivalent to that derived by \citet{mihalas:1978} in the limit that the interior radiation term (i.e., infrared radiation emitted by the surface) dominates the shortwave stellar radiation directly absorbed by the atmosphere.  Note that at higher pressures ($>10^6$ Pa), the assumption that the atmosphere is transparent to short wavelengths may not be physically valid, and the lower levels of these thicker atmospheres may be optically thick at short wavelengths.

Note that, within the framework of our model, it is possible to rewrite $\tau$ as $\tau= \tau_{oo}(p/p_{oo})^n$, where $n=2$ represents a source of collision-induced opacity~\citep{lorenz:2003}.  \citet{merlis:2010} include an additional term that is proportional to the fourth power of pressure.  \citet{heng:2012b} have developed a more complicated semi-analytical formalism for calculating the temperature–-pressure profiles in hot Jupiter atmospheres with a correction factor for collision-induced opacity.  Collision-induced opacity is an avenue of future improvement to our model.

	In the mathematical limit $\tau \rightarrow \tau_s$ in Eq.~\ref{eq:teq}, there is a temperature discontinuity between the pure radiative equilibrium temperature of the surface and the pure radiative equilibrium temperature of the atmosphere just above the surface.  To mitigate this unphysical property of the solution, convective adjustment is performed~\citep[see][for details of this scheme]{lorenz:2003,zalucha:2010} on the pure radiative equilibrium atmospheric and surface temperatures.  In radiative-convective equilibrium, the temperature profile within the near-surface convective layer follows an adiabat and is continuous with the surface temperature.  Explicitly, the radiative-convective equilibrium temperature is
	\begin{equation}
	\sigma T^4_{eq,c}=\left(1-A\right)Q_o\left(0.5+0.75\tau_t\right)\left[\frac{\tau}{\tau_t}\right]^{4R/c_p}
	\end{equation}
for $\tau \geq \tau_t$, where $\tau_t$ is the $\tau$ at the top of the convective layer.  For $\tau < \tau_t$, $T_{eq,c}$ is given by Eq.~\ref{eq:teq}.  $\tau_t$ is solved by numerically balancing the upward surface flux in pure radiative equilibrium, the integrated upward atmospheric flux for $\tau \geq \tau_t$ in pure radiative equilibrium, the upward surface flux in radiative-convective equilibrium, and the integrated upward atmospheric flux in the convective layer in radiative-convective equilibrium.  Specifically, 
\begin{eqnarray}
\lefteqn{\sigma T_{eq,s}^4 \exp\left[-\left(\tau_s - \tau_t\right)/\mu\right] + \int^{\tau_s}_{\tau_t} \sigma T_{eq}^4(\tau_*) \exp\left[-\left(\tau_* - \tau_t\right)/\mu\right] d\tau_*} \label{eq:fluxbalance} \\
& & =\sigma T_{eq,c}^4(\tau_s) \exp\left[-\left(\tau_s - \tau_t\right)/\mu\right] + \int^{\tau_o}_{\tau_t} \sigma T_{eq,c}^4(\tau_*) \exp\left[-\left(\tau_* - \tau_t\right)/\mu\right] d\tau_* \nonumber
\end{eqnarray}
where $\mu=2/3$ is the average cosine of the emission angle and $\tau_*$ is a variable of integration.  $T_{eq,c}$ is now quantitatively known for use in Eq.~\ref{eq:newtonian}.

  Under the condition of surface-atmosphere mass exchange, the atmosphere deposits ice on the surface (if the atmospheric temperature drops below the condensation temperature) or the surface sublimates ice into the atmosphere (if the surface temperature is greater than the condensation temperature).  In both cases the atmosphere remains at the condensation temperature.  For arguments of simplicity, we do not include surface-atmosphere mass exchange (but this is an avenue of future work).  We do account for the temperature exchange by prescribing that if the radiative equilibrium or GCM-predicted temperature drops below the condensation temperature (such as on the nightside of the planet where the insolation is zero), the temperature is instantaneously reset to the condensation temperature.  Otherwise, the temperature could drop to absolute zero, which is unphysical.	This convention has been demonstrated to be valid in the Mars community~\citep{haberle:1993,takahashi:2003,zalucha:2010}, except near the poles where surface-atmosphere mass exchange occurs.  This step implicitly supplies the latent heat that would have been added/subtracted, thus conserving energy. The condensation temperature is given by the Clausius--Clapeyron relation
	\begin{equation}\label{eq:cc}
	T_{cond}=\left[\frac{1}{T_1}-\frac{R}{L}\ln\left(\frac{p}{p_1}\right)\right]^{-1},
	\end{equation}
	where $T_1$ and $p_1$ are the vaporization reference temperature and pressure, respectively (see Table~\ref{tb:gcmall}) and $L$ is the specific latent heat of vaporization.  Note that $L$ has a dependence on temperature, which we do not take into account (since we are not aware of any measurements taken at the high temperatures encountered on hot exoplanets).

	\section{Simulation results}\label{se:results}
\begin{figure}
 \noindent\includegraphics[width=0.8\textwidth,angle=270]{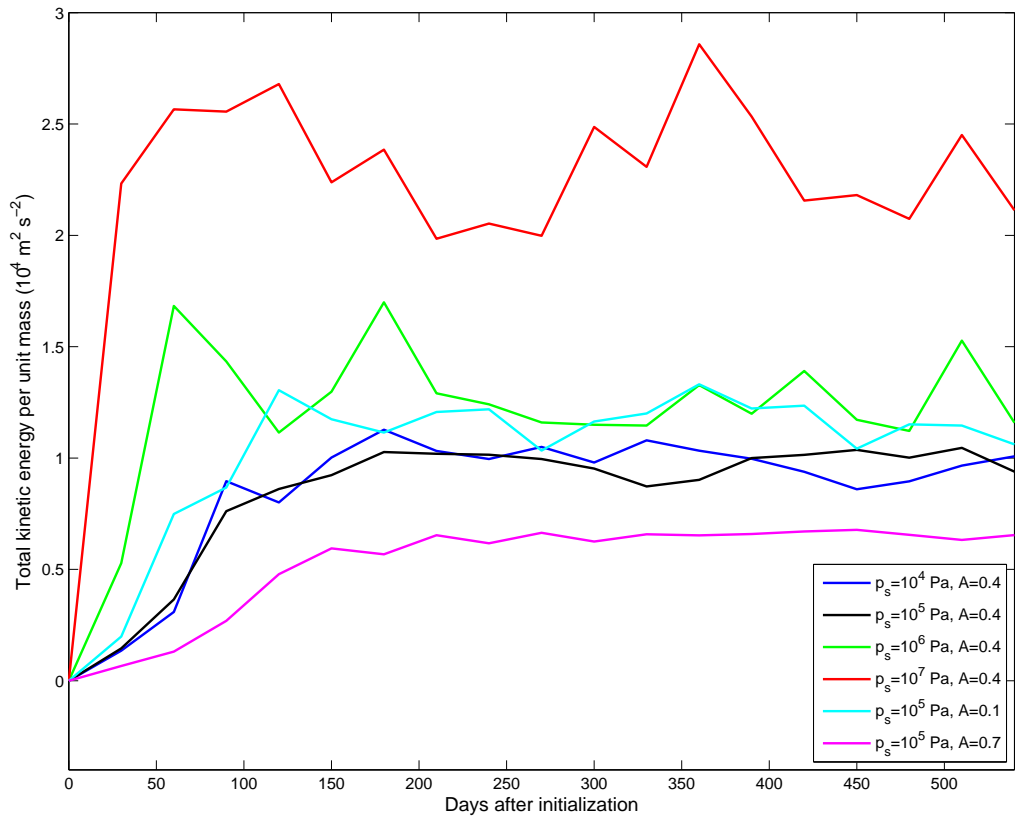}
 \caption{\label{fg:kespinup} Globally averaged kinetic energy per unit mass from select GCM runs.  The kinetic energy begins from a rest state (zero) and increases as the model spins up.  The level of kinetic energy then fluctuates due to statistical variations in the model, but maintains a flat trend.  The results presented subsequently in this work are taken from this flat period.}
 \end{figure}

	We performed simulations starting from rest, initializing with a globally constant temperature and surface pressure.  Figure~\ref{fg:kespinup} shows the globally averaged kinetic energy per unit mass for selected cases~\citep[see e.g.][for a demonstration of this method]{leovy:1969}.  After 120 Earth-days, the model reached a quasi-steady state.  The results presented for the remainder of this paper are the time average of the subsequent 400~Earth-days.  A 120 Earth-day spin up time is short compared to the 200 Earth-day spin up time of the benchmark \citet{held:1994} Earth model.  We suspect that this is because our radiative relaxation time is shorter than their 40 Earth-day value.  
	
	The globally averaged surface pressure, which is a representation of the atmosphere's mass, is an important but unconstrained parameter for super-Earths.  Since our atmosphere is assumed to be entirely made up of the radiatively active constituent, the global mean surface pressure is directly tied to the strength of the greenhouse effect.  Higher global mean surface pressure means more IR absorption, and thus higher temperatures.  We have chosen to perform simulations using the following global mean surface pressures: $10^4$, $10^5$, $10^6$, and $10^7$~Pa.  For reference, the global mean surface pressures of Venus, Earth, Mars, and Titan are approximately $90 \times 10^5$~Pa, $10^5$~Pa, $0.006 \times 10^5$~Pa, and $1.5 \times 10^5$~Pa, respectively; however, note that none of these atmospheres are purely greenhouse-gas dominated as we have assumed in our study.
	
	The geometric albedo of a planet is also an important parameter for planetary atmospheres because it controls the amount of energy that is absorbed by the planet's cross-section, and thus the amount of energy absorbed by the surface and re-emitted into the atmosphere.  Albedo variations (e.g. due to inhomogeneous surface distributions of oceans, rocks, and/or ices) over the planet's surface can also be important.  For simplicity, we perform simulations with globally constant surface albedos of 0.1, 0.4, and 0.7.  Note that the globally averaged surface albedos of Venus, Earth, Mars, and Titan are approximately 0.77, 0.3, 0.15, and 0.22, respectively.

	\subsection{Results of the $10^5$~Pa surface pressure, 0.4 surface albedo case}
		
	We examine in detail the results of the $10^5$~Pa global mean surface pressure and 0.4 surface albedo case.  This case is near the center of our parameter sweep and is thus a prudent starting point for our discussion.
	
\begin{figure}
 \noindent\includegraphics[width=0.8\textwidth,angle=270]{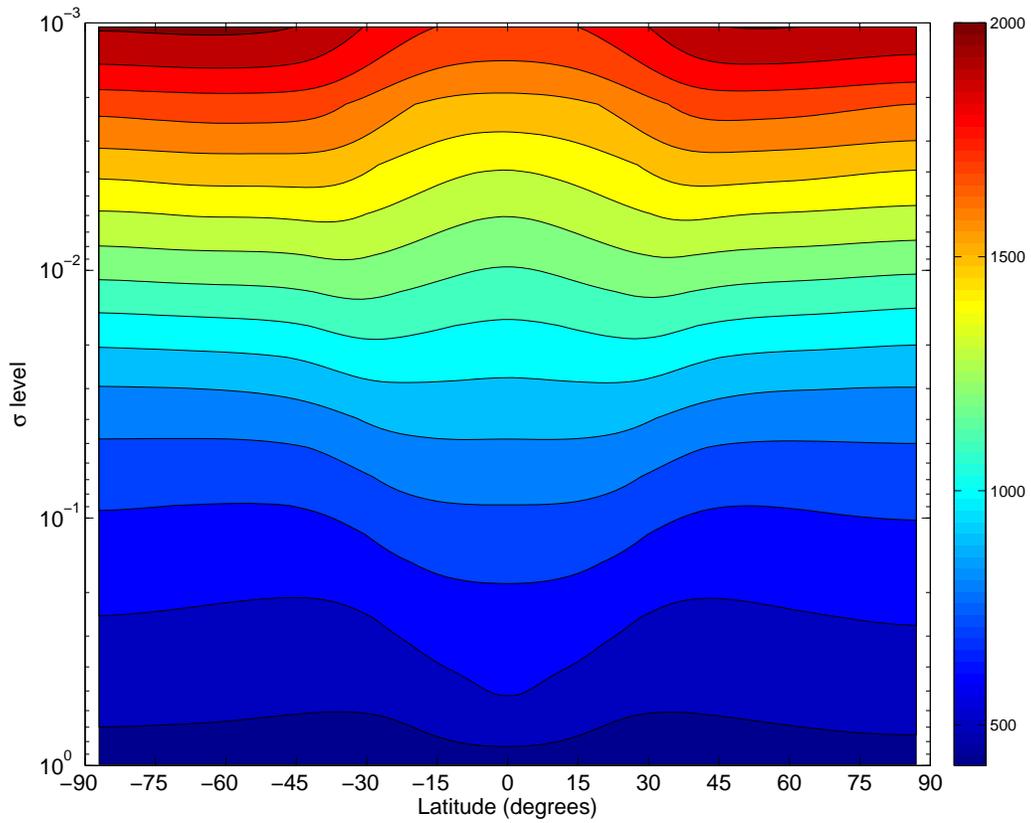}
 \caption{\label{fg:thetamean} Zonally averaged potential temperature (K) from GCM (global mean surface pressure=$10^5$~Pa, surface albedo=0.4 case).  Contour interval is 100~K.  Time averaging was performed over 400 Earth-days.  }
 \end{figure}

		Figure~\ref{fg:thetamean} shows the zonally averaged potential temperature.  This plot and the subsequent zonal mean plots are plotted with $\sigma$ level $=p/p_s$ as the vertical coordinate (i.e., the plots have been normalized to the global mean surface pressure).  At $\sigma$ levels greater than 0.02 (pressure $2\times 10^3$~Pa), at a fixed level, there is a dip in potential temperature in the tropics.  At lower $\sigma$ levels, there is a rise in potential temperature in the tropics.

\begin{figure}
 \noindent\includegraphics[width=0.8\textwidth,angle=270]{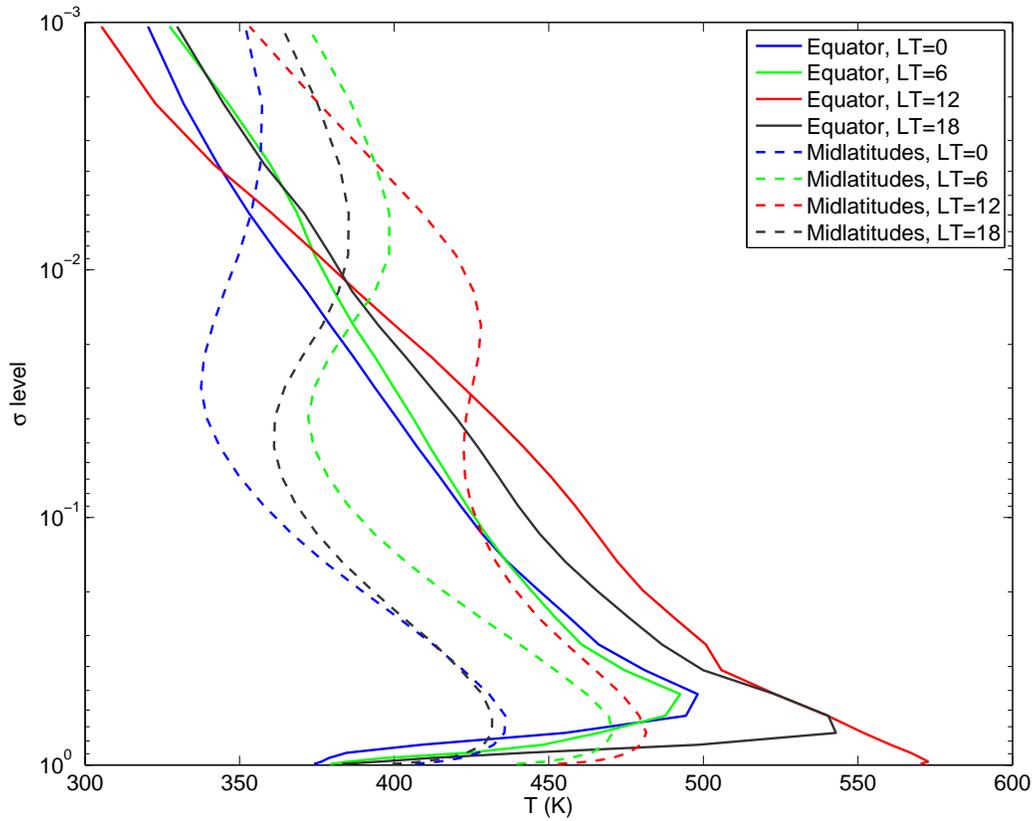}
 \caption{\label{fg:tltandlat} Temperature profiles from GCM (global mean surface pressure=$10^5$~Pa, surface albedo=0.4 case).  The profiles are time averaged over 400 Earth-days.  Two latitudes are shown: the equator and 60\de (midlatitudes).   Local times (LT) are plotted for the nightside (0 LT), dayside (12 LT), and morning (6 LT) and evening (18 LT) terminators.}
 \end{figure}

	Figure~\ref{fg:tltandlat} shows temperature profiles from the equator and 60\de~latitude (midlatitude) at four local times.  All of the temperature profiles contain a temperature inversion (temperature increasing with decreasing pressure) near the surface, and the midlatitude profiles contain another inversion peaking at about the 0.01 $\sigma$ level (pressure $10^3$ Pa).  The near-surface inversions are also present in some of the other global mean surface pressure and surface albedo cases.  The precise cause of this inversion is unknown.  The underlying radiative-convective temperature monotonically decreases with decreasing pressure (Eq.~\ref{eq:teq}) because there are no assumed shortwave absorbers and the optical depth decreases linearly with pressure.  Moreover, on the nightside, where the insolation is zero (and therefore the radiative-convective temperature would be unphysically at absolute zero), the column radiative-convective temperature is set to the condensation temperature.  Since the radiative relaxation timescale is assumed constant with height, it is puzzling that the surface temperature would be pulled more strongly towards colder radiative-convective equilibrium temperature and yet at a slightly higher altitude, less so.
	
	  Temperature inversions may be caused dynamically be two physical processes: downwelling, which leads to adiabatic compression/heating, and differential horizontal heat transport with height.  We have searched through all of our cases and cannot find a direct correlation between the presence of an inversion and downwelling.  We have not analyzed  the local horizontal heat transport, as this is beyond the scope of this paper.  We speculate that the cause of the temperature inversion is some complex combination of these two physical processes.
	
	Comparing pressure-temperature ($p$-$T$) profiles at the same local times indicates that at some altitudes the temperature is higher in the midlatitudes, which would not be expected through astronomical considerations alone because the stellar flux decreases as a function of latitude.  The reversal occurs at pressures lower than about the 0.01 $\sigma$ level ($10^3$ Pa) and within a few $\sigma$ level tenths ($10^4$ Pa) of the surface.  There is also a longitudinal asymmetry between the morning and evening terminators.  At the equator, the evening temperature profile is greater than or equal to the morning temperature at all levels, while in the midlatitudes, the opposite is the case.  Note that the temperature profiles are always greater than the condensation temperature of \htwoo, because we have explicitly forbidden it in the model equations and parameterizations.
	
\begin{figure}
 \noindent\includegraphics[width=1.0\textwidth]{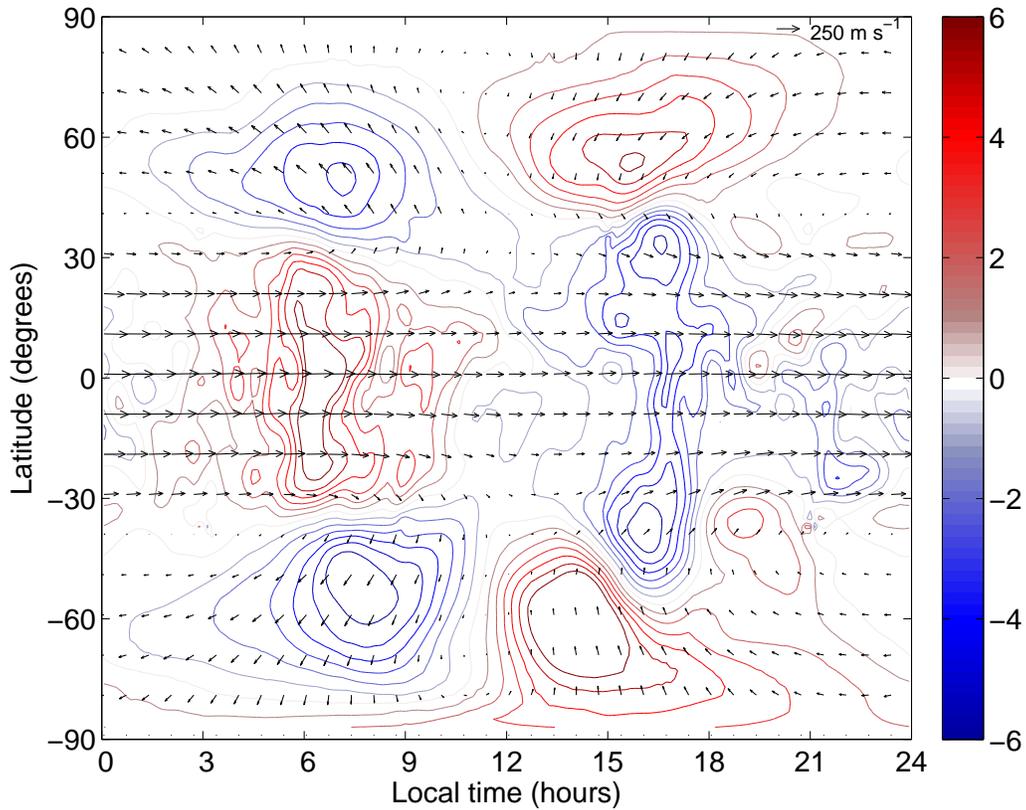}
 \caption{\label{fg:quiver16} Horizontal winds (arrows, in m~s$^{-1}$) and vertical winds in pressure coordinates (colored contours, in increments of $10^{-2}$~Pa~s) from GCM simulation with global mean surface pressure=$10^5$~Pa and surface albedo=0.4.  Time averaging was performed over 400 Earth-days.  The winds are shown as an example of the circulation in the free atmosphere at $\sigma$ level 0.068 (pressure 6800~Pa).  Note that in pressure coordinates positive vertical wind is downward.}
 \end{figure}

	Figure~\ref{fg:quiver16} shows the wind structure of the free atmosphere (above the frictional boundary layer) at $\sigma$ level 0.068 (pressure 6800~Pa).  While quantitative differences exist at the levels outside the boundary layer (pressures less than $0.7~\times 10^5$~Pa), they all exhibit a stationary wavenumber-1 longitudinal structure in the vertical velocity field.  The circulation is divided into two latitudinal regimes.  First, an equatorial region between $\pm30^{\circ}$ with a high speed ($\ge 300$~m~s$^{-1}$) westerly jet\footnote{For reference, the sound speed in an ideal gas of water vapor is 495~m~s$^{-1}$ at 400~K and 858~m~s$^{-1}$ at 1200~K}, sinking motion on the western (i.e., morning) hemisphere, and rising motion in the eastern (i.e., evening) hemisphere.  Second, poleward of $\pm35$\de~latitude, the flow is northward over the pole on the terminators and dayside and eastward around the nightside of the planet.  Moreover, the sign of the vertical motions switch such that descent is on the evening hemisphere and ascent is on the morning hemisphere.
	
\begin{figure}
 \noindent\includegraphics[width=0.8\textwidth,angle=270]{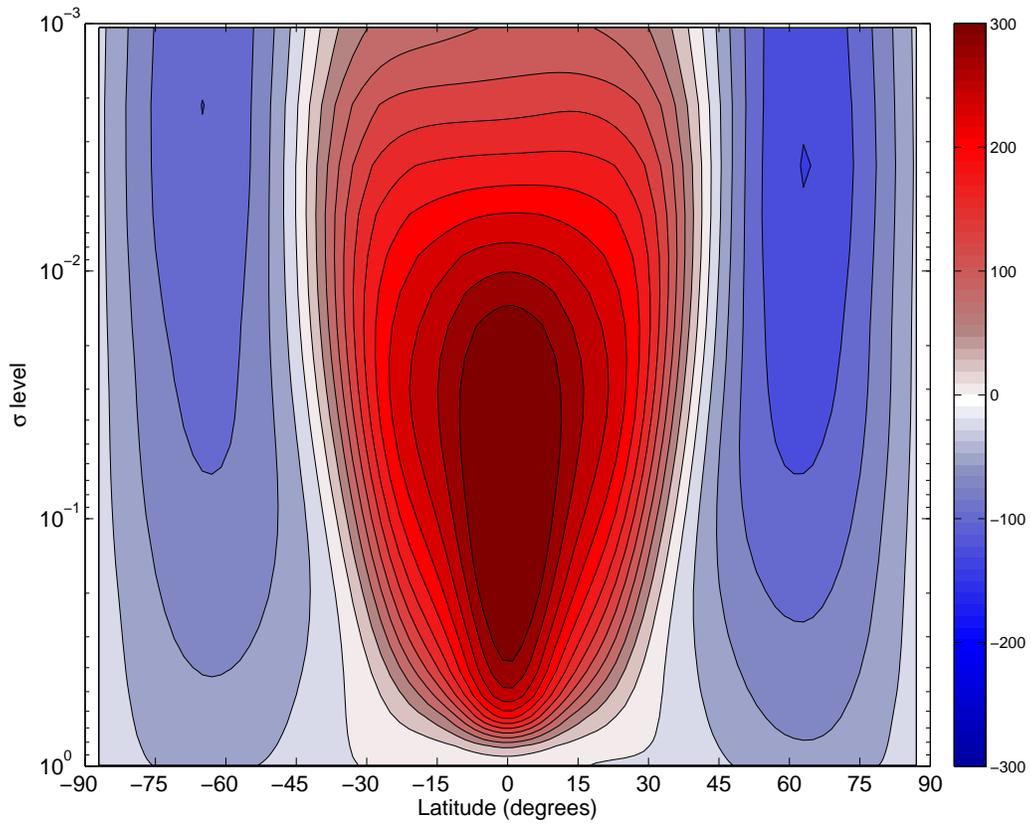}
 \caption{\label{fg:umean} Zonally averaged zonal winds (\ms) from GCM (global mean surface pressure=$10^5$~Pa, surface albedo=0.4 case).  Contour interval is 25~\ms.  Time averaging was performed over 400 Earth-days.  The westerly equatorial jet has a width of about 30--45\de. }
 \end{figure}

\begin{figure}
 \noindent\includegraphics[width=0.8\textwidth,angle=270]{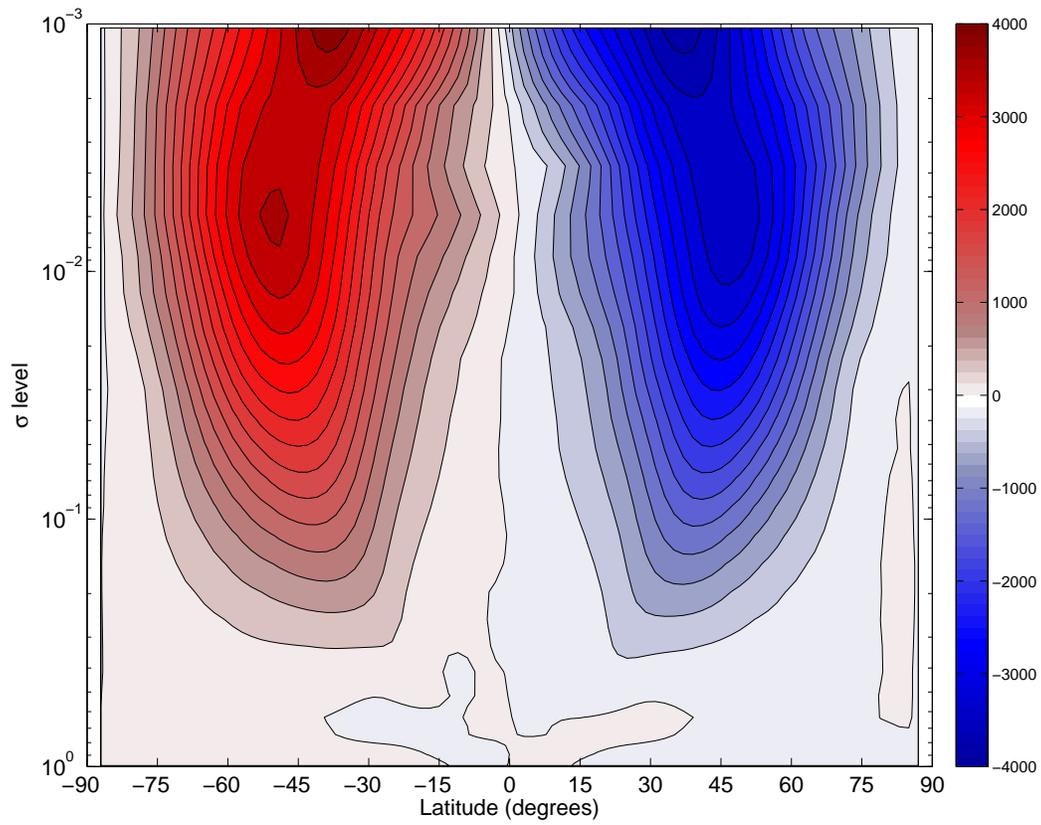}
 \caption{\label{fg:etmean} Latitudinal transport of eastward eddy momentum per unit mass (m$^2$~s$^{-2}$) from GCM (global mean surface pressure=$10^5$~Pa, surface albedo=0.4 case).  Contour interval is 250~m$^2$~s$^{-2}$.  Time averaging was performed over 400 Earth-days.  }
 \end{figure}

Figure~\ref{fg:umean} shows the zonally averaged zonal mean wind.  There is a slight dependence on altitude, but westerly winds occur in the tropics between $30$\de~to 40\de~in each hemisphere.  Easterly winds occur at higher latitudes.  The maximum westerly wind strength is achieved at around $\sigma$ level 0.1 (pressure $10^4$~Pa), while the maximum easterly wind strength occurs at around $\sigma$ level 0.002 (pressure 200~Pa).  Figure~\ref{fg:etmean} shows the latitudinal transport of westerly eddy momentum (per unit mass), $\overline{u'v'}$, where $u$ is the zonal velocity, $v$ is the meridional velocity, primes denote deviations from the zonal means, and the overbar denotes a zonal average.  This figure shows that in the Northern hemisphere, $\overline{u'v'} < 0$, indicating southward transport of westerly momentum, while in the Southern hemisphere $\overline{u'v'} > 0$, indicating northward transport.  Thus, westerly eddy momentum is converging at the equator.  The magnitude of the transport increases with height at all latitudes, and is strongest at approximately $\pm$45\de.

\begin{figure}
 \noindent\includegraphics[width=0.8\textwidth,angle=270]{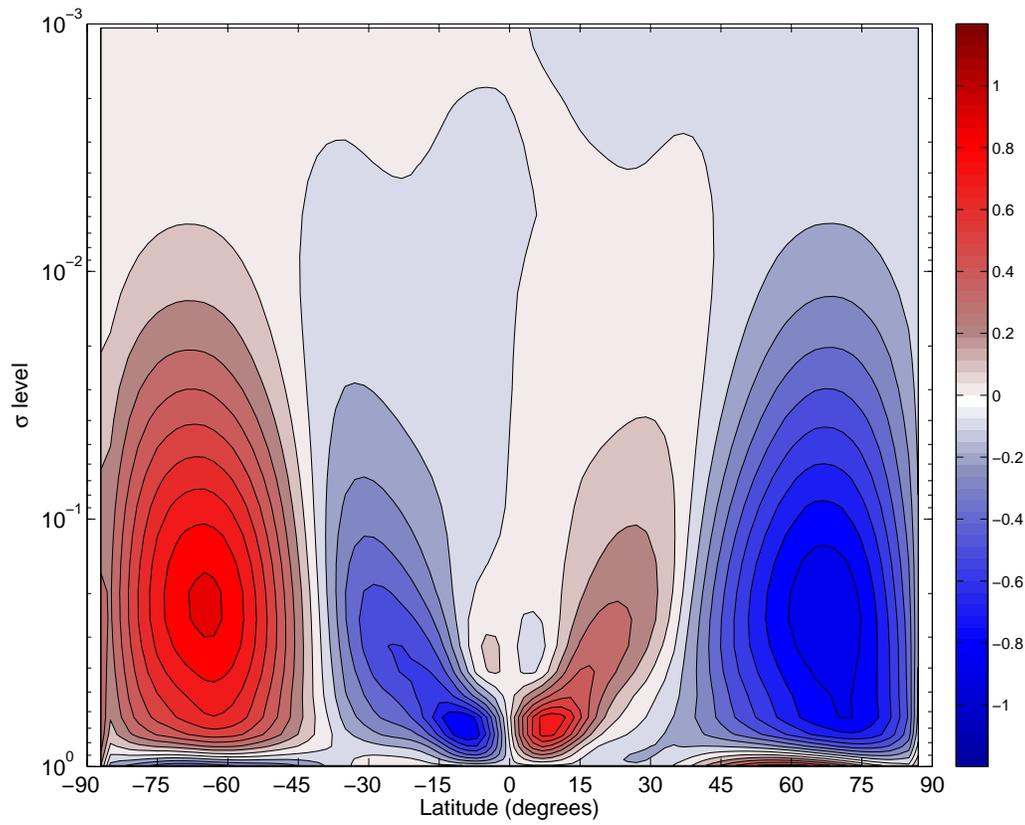}
 \caption{\label{fg:psimean} Zonally averaged mass stream function ($10^{12}$~kg~s$^{-1}$) from GCM (global mean surface pressure=$10^5$~Pa, surface albedo=0.4 case).  Contour interval is $0.1\times 10^{12}$~kg~s$^{-1}$.  Time averaging was performed over 400 Earth-days.  Positive flow is clockwise.}
 \end{figure}

The two latitudinal regimes seen in the zonal wind fields may also be seen in a plot of zonally averaged Euler-mean mass stream function (Fig.~\ref{fg:psimean}).  In each hemisphere, there is a cell that rises at the equator, flows poleward aloft, and sinks at $\sim 35$\de~latitude.  Poleward of this latitude, there is a cell that circulates in the opposite sense and is stronger than the equatorial flows.  The one Earth-day rotation period case of \citet{merlis:2010} shows a three celled structure.  Moreover, their maximum stream function magnitude is two orders of magnitude lower than ours.  This is consistent with our higher temperatures and zonal and meridional wind speeds.  The atmosphere is more energetic and must transport heat away from the substellar point more vigorously.
	
	In the boundary layer at $\sigma$ level 0.9377 (pressure 93770~Pa), the circulation has a different structure than in the free atmosphere.  The midlatitudes are characterized by weak downwelling and a horizontal wavenumber-1 pattern.  At the equator, an intense area of downwelling and easterly flow is present at the substellar point (12 hours LT), while there is an area of intense upwelling at 15 hours LT.  Still farther to the east the upward motion switches back to downward.
		
	\subsection{Results of differing surface pressure and surface albedo}		
	
	The global mean surface pressure (i.e. the atmosphere's total mass) and the surface albedo are not well constrained by observations; however, these are fundamental parameters that strongly modulate atmospheric processes.  In order to compare simulations with different global mean surface pressures, subsequent quantities are plotted on the same $\sigma$ level .  If we instead chose a constant pressure level, we might be comparing a level that is in the boundary layer of a low surface pressure case with a level that is near the model top in a high pressure case, where different physical processes are acting.
	
\begin{figure}
 \noindent\includegraphics[width=0.8\textwidth,angle=270]{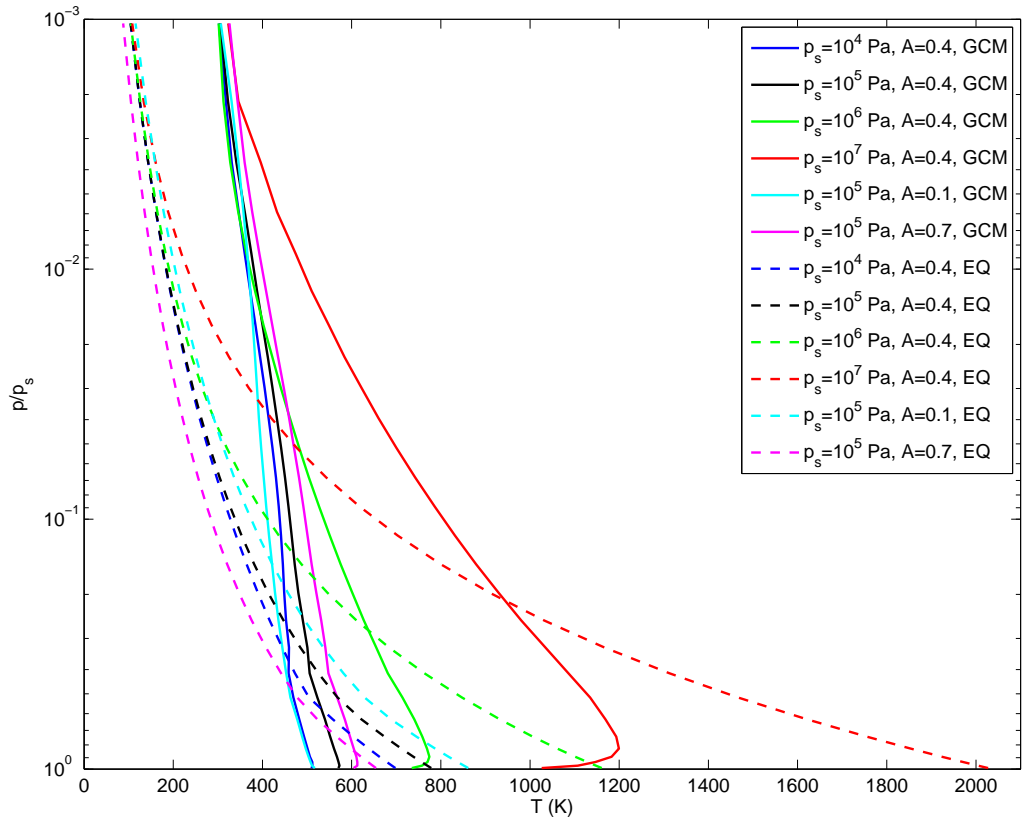}
 \caption{\label{fg:tpsanda} Temperature profiles as a function of global mean surface pressure and surface albedo.  GCM refers to the dynamical temperatures output from the GCM, and EQ refers to the radiative-convective equilibrium temperature.  The profiles are time averaged over 400 Earth-days.  All profiles are shown for the equator and 12 hours LT.  }
 \end{figure}

	Figure~\ref{fg:tpsanda} shows dayside, equatorial temperature profiles as a function of $p_s$ and $A$ for both the GCM simulated temperatures and the radiative-convective equilibrium temperatures (see Section~\ref{ss:rs}).  As $p_s$ increases, temperature increases.  This behavior occurs simply because we have assumed that the atmosphere is entirely composed of the IR absorber, so as atmospheric mass increases, so does the amount of the absorber.  The atmosphere subsequently absorbs more of the energy being reradiated by the surface.  Similarly, temperature decreases as $A$ increases because more of the incoming stellar flux is reflected to space.  In general, the GCM simulated temperatures are cooler than the radiative-convective equilibrium temperatures (for a given $p_s$ and $A$) in the lower atmosphere, but warmer in the upper atmosphere.  The GCM temperatures are more isothermal.  By definition, the radiative equilibrium temperature is the temperature of a motionless atmosphere, so any difference is indicative of circulation.  In Fig.~\ref{fg:tpsanda}, adiabatic heating due to downward motion is either occurring on the precisely at LT=12 hours and the equator, or at another location with subsequent lateral transport to LT=12 hours at the equator.

	These may be lateral heat transport by horizontal winds and/or adiabatic heating and cooling due to downward and upward motions, respectively.

\begin{figure}
 \noindent\includegraphics[width=0.9\textwidth,angle=270]{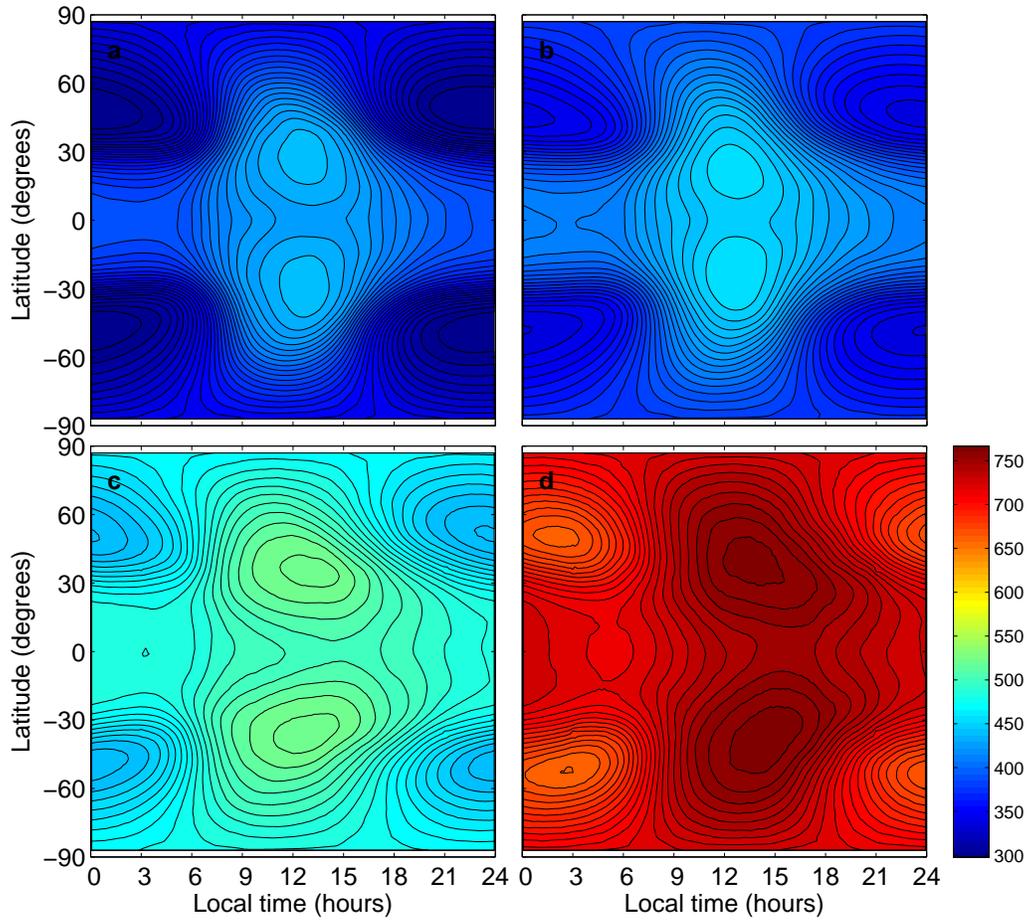}
 \caption{\label{fg:tps16} GCM temperature (K) as a function of global mean surface pressure: (a) $p_s=10^4$~Pa, (b) $p_s=10^5$~Pa, (c) $p_s=10^6$~Pa, and (d) $p_s=10^7$~Pa.  All cases have $A=0.4$, are time averaged over 400 Earth-days, and are shown at the same $\sigma$ level $0.068$.  Contour interval is 5~K.}
 \end{figure}

	Figure~\ref{fg:tps16} shows GCM temperature in the free atmosphere (outside of the frictional boundary layer) as a function of $p_s$ with $A$ held constant.  As $p_s$ increases, the temperature increases at all locations, for reasons described above.  In all cases, the equatorial temperature maximum is shifted eastward of the substellar point, as is also the case in hot Jupiter models~\citep[e.g.,][]{showman:2008,perna:2012}.  The maximum temperature exists offset from the equator at about 30\de~latitude and 12 hours LT.  Thus, on the dayside as we move from equator to pole, the temperature first increases until this location is reached, then decreases towards the poles.  At high latitudes, the temperature on a given latitude circle is maximum near 12 hours LT, but the exact longitude of the maximum is located on either side of this meridian depending on the direction of the zonal flow.  A temperature minimum occurs at 0 hours LT and 50\de~latitude.  On the nightside, the temperature decreases monotonically from equator to pole. 
	
\begin{figure}
 \noindent\includegraphics[width=1.0\textwidth,angle=270]{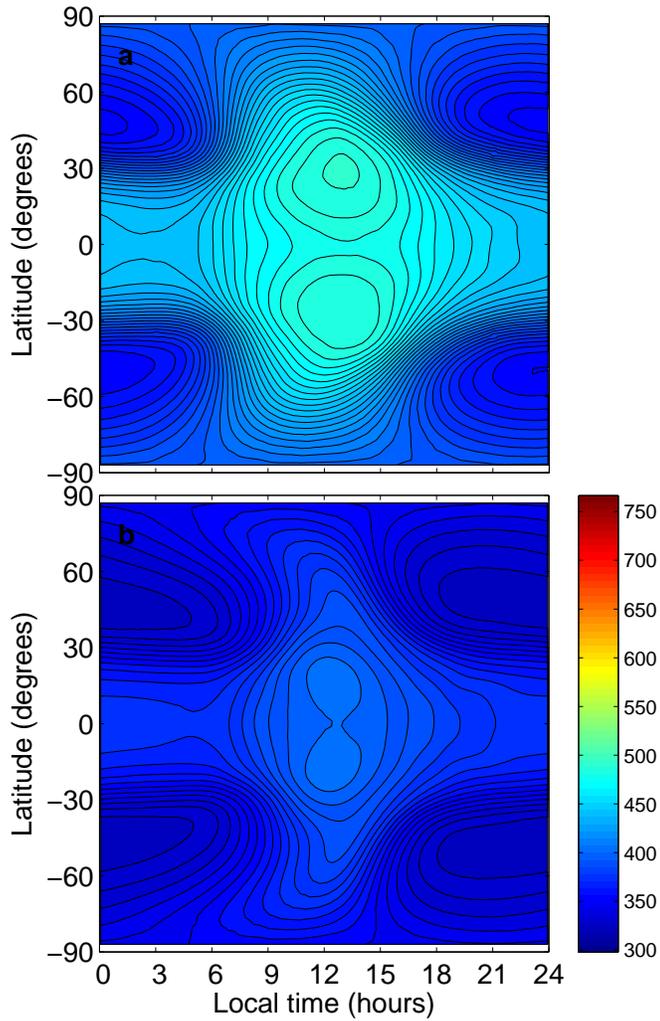}
 \caption{\label{fg:ta16} GCM temperature (K) as a function of surface albedo: (a) $A=0.1$, (b) $A=0.7$.  All cases have $p_s=10^5$~Pa, are time averaged over 400 Earth-days, and are shown at the same $\sigma$ level $0.068$.  Contour interval is 5~K.  Taken with panel (b) of Fig.~\ref{fg:tps16}, these plots form a progression of surface albedos from 0.1--0.7.  }
 \end{figure}

Figure~\ref{fg:ta16}, along with panel (b) of Fig.~\ref{fg:tps16} (in the order Fig.~\ref{fg:ta16}a, Fig.~\ref{fg:tps16}b, Fig.~\ref{fg:ta16}b), show the effect of increasing $A$ on temperature in the free atmosphere ($p_s$ held constant).  Again, the effect on the horizontal structure is simply to decrease the temperature everywhere as $A$ increases.

\begin{figure}
 \noindent\includegraphics[width=0.9\textwidth,angle=270]{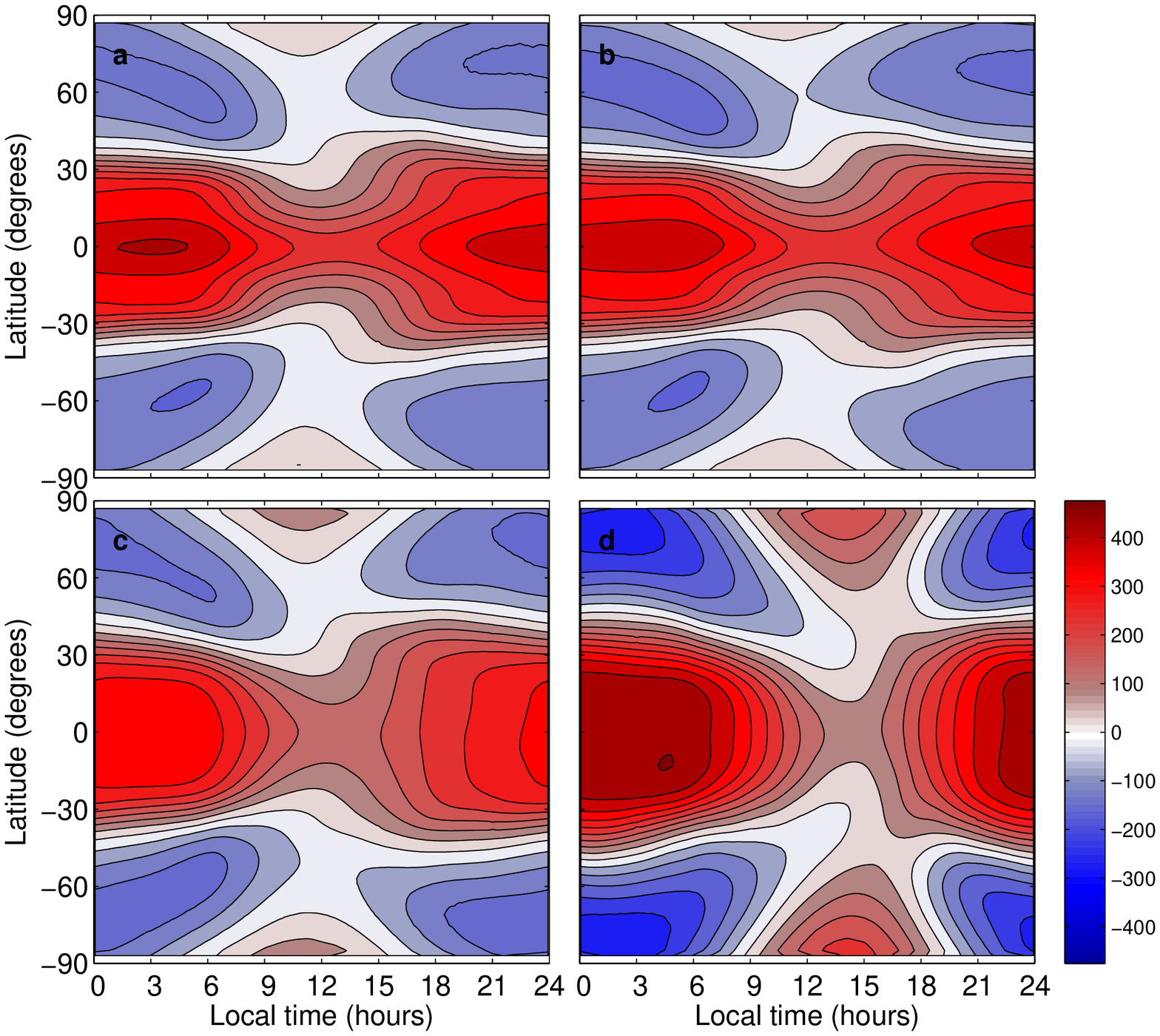}
 \caption{\label{fg:ups16} GCM zonal winds (\ms) as a function of global mean surface pressure: (a) $p_s=10^4$~Pa, (b) $p_s=10^5$~Pa, (c) $p_s=10^6$~Pa, and (d) $p_s=10^7$~Pa.  All cases have $A=0.4$, are time averaged over 400 Earth-days, and are shown at the same $\sigma$ level $0.068$.  Contour interval is 50~\ms.  Positive corresponds to westerly flow.}
 \end{figure}

\begin{figure}
 \noindent\includegraphics[width=1.0\textwidth,angle=270]{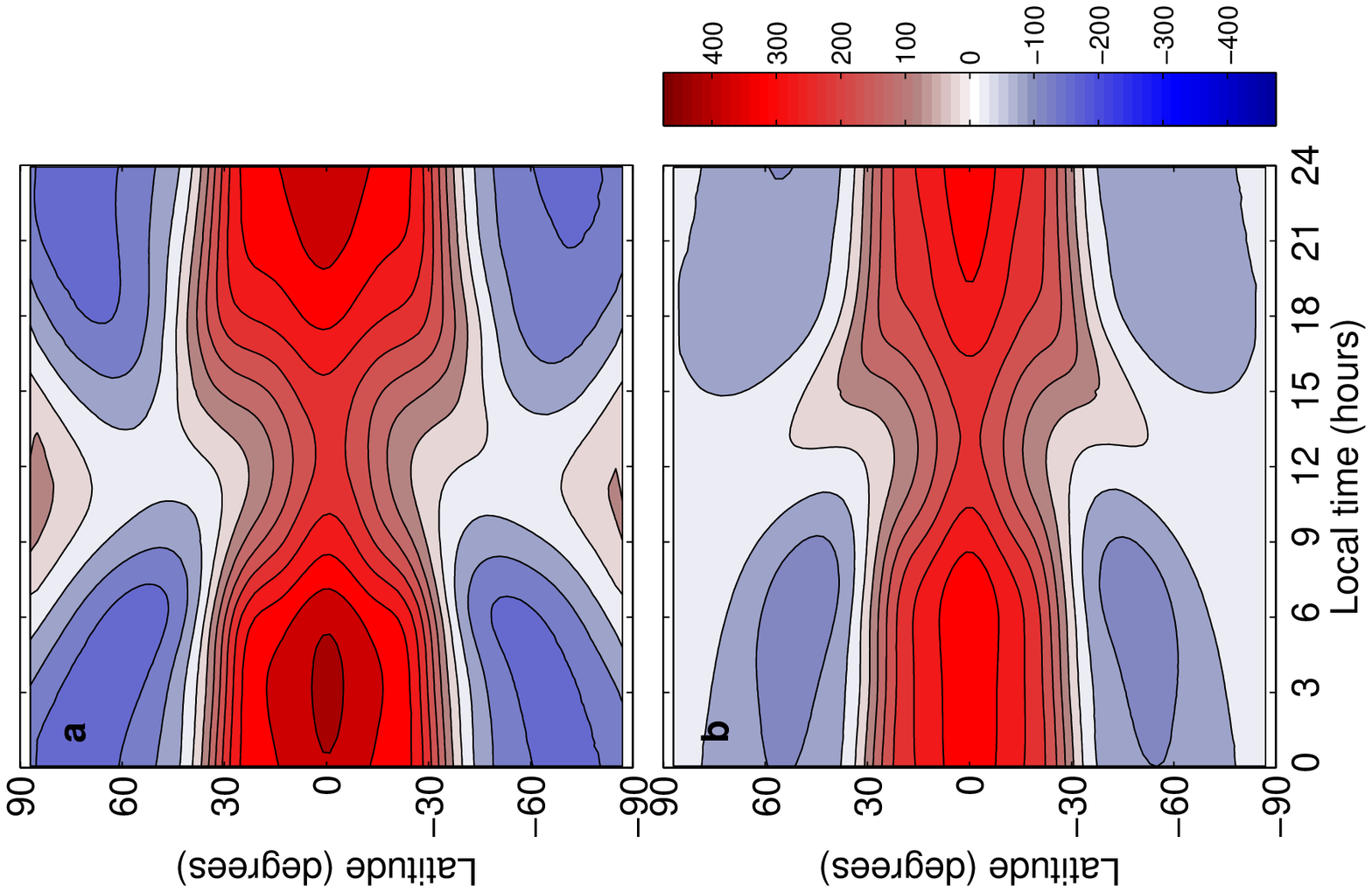}
 \caption{\label{fg:ua16} GCM zonal winds (\ms) as a function of surface albedo: (a) $A=0.1$, (b) $A=0.7$.  All cases have $p_s=10^5$~Pa, are time averaged over 400 Earth-days, and are shown at the same $\sigma$ level $0.068$.  Contour interval is 50~\ms.  Taken with panel (b) of Fig.~\ref{fg:ups16}, these plots form a progression of surface albedos from 0.1--0.7.  Positive corresponds to westerly flow.}
 \end{figure}

Figure~\ref{fg:ups16} shows the zonal component of wind in the free atmosphere as a function of $p_s$.  As $p_s$ increases, the speed of the equatorial jet decreases between 9 and 15 hours LT.  The latitudinal extent of the equatorial jet increases on the nightside.  At the two greatest initial surface pressure cases ($10^6$ and $10^7$~Pa), an increasingly westerly component of flow develops at middle to high latitudes between 11 and 18 hours LT.  The nightside, easterly flow in the midlatitudes also increases in speed as $p_s$ increases.  Figure~\ref{fg:ua16}, along with panel (b) of Fig.~\ref{fg:ups16} (in the order Fig.~\ref{fg:ua16}a, Fig.~\ref{fg:ups16}b, Fig.~\ref{fg:ua16}b), show the effect of increasing surface albedo on zonal wind in the free atmosphere.  The zonal wind speed in the equatorial and midlatitude jets decreases with increasing $A$.  Also, the latitudinal width of the equatorial jet decreases as $A$ increases.

\begin{figure}
 \noindent\includegraphics[width=0.9\textwidth,angle=270]{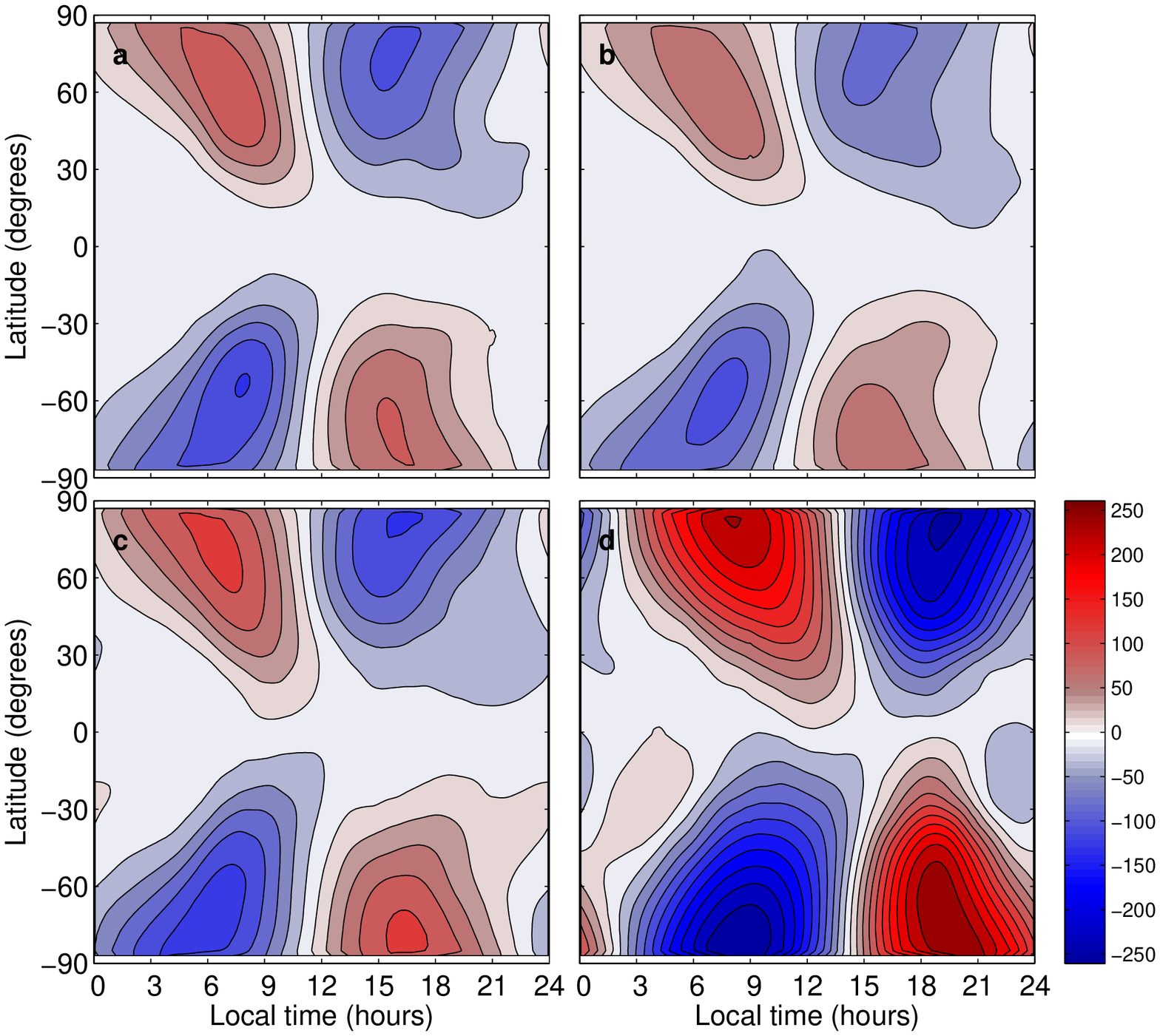}
 \caption{\label{fg:vps16} GCM meridional winds (\ms) as a function of global mean surface pressure: (a) $p_s=10^4$~Pa, (b) $p_s=10^5$~Pa, (c) $p_s=10^6$~Pa, and (d) $p_s=10^7$~Pa.  All cases have $A=0.4$, are time averaged over 400 Earth-days, and are shown at the same $\sigma$ level $0.068$.  Contour interval is 25~\ms.  Positive corresponds to southerly flow.}
 \end{figure}

\begin{figure}
 \noindent\includegraphics[width=1.0\textwidth,angle=270]{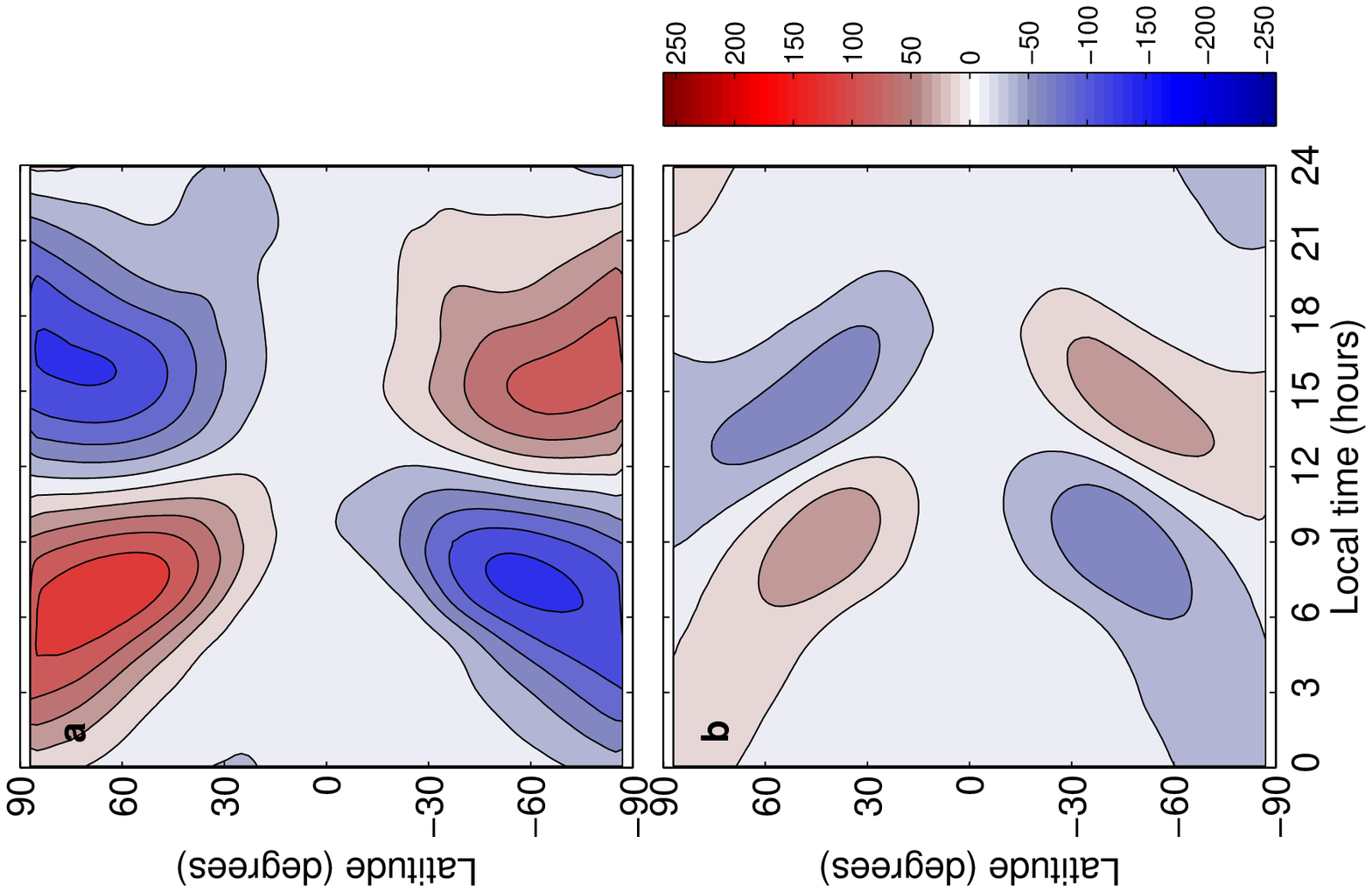}
 \caption{\label{fg:va16} GCM meridional winds (\ms) as a function of surface albedo: (a) $A=0.1$, (b) $A=0.7$.  All cases have $p_s=10^5$~Pa, are time averaged over 400 Earth-days, and are shown at the same $\sigma$ level $0.068$.  Contour interval is 25~\ms.  Taken with panel (b) of Fig.~\ref{fg:vps16}, these plots form a progression of surface albedos from 0.1--0.7.  Positive corresponds to southerly flow.}
 \end{figure}

Figures~\ref{fg:vps16} and \ref{fg:va16} show the effect of increasing $p_s$ and $A$, respectively, on meridional flow in the same order as Figs.~\ref{fg:ups16} and \ref{fg:ua16}.  All cases have areas of southerly flow at southern longitudes from 12--24 hours LT and northern longitudes from 0--12 hours LT and northerly flow elsewhere.  In the cylindrical projection format of our plots, this behavior indicates that there is a component of the flow over the poles from the evening side (i.e., western hemisphere) to the morning side (i.e., eastern hemisphere).  As $p_s$ increases, the meridional wind speed increases; as $A$ increases the speed decreases.  This behavior implies a link between the amount of thermal energy and the strength of the winds.

\begin{figure}
 \noindent\includegraphics[width=0.9\textwidth,angle=270]{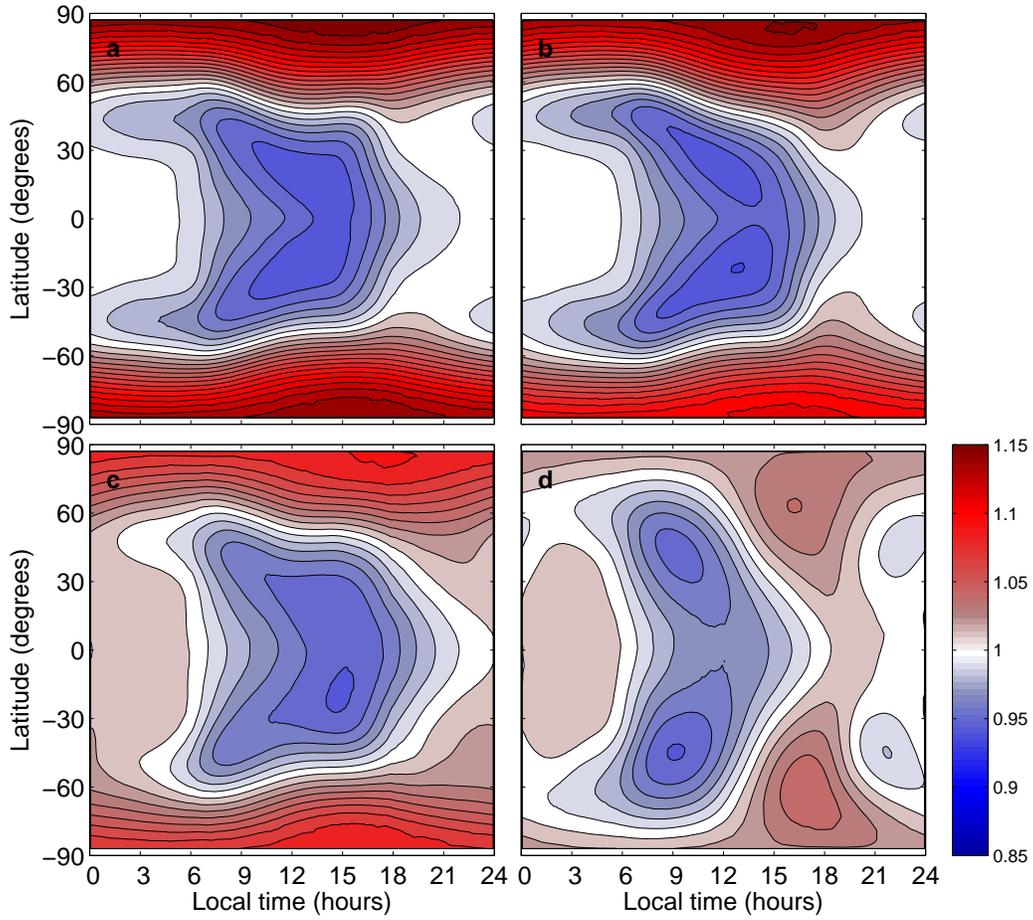}
 \caption{\label{fg:psps} GCM surface pressure (Pa) variation as a function of global mean surface pressure: (a) $p_s=10^4$~Pa, (b) $p_s=10^5$~Pa, (c) $p_s=10^6$~Pa, and (d) $p_s=10^7$~Pa.  Each panel is normalized by the global mean surface pressure for that case.  Contour interval is 1\% of the global mean surface pressure.  All cases have $A=0.4$ and are time averaged over 400 Earth-days.}
 \end{figure}

	Figure~\ref{fg:psps} shows the time-integrated surface pressure as a function of latitude and longitude and $p_s$.  Note the distinction between the global mean surface pressure ($p_s$) and the latitudinally and longitudinally varying surface pressure---the latter is equal to $p_s$ in the global sum.  To show each case on the same scale, the latitudinally and longitudinally varying surface pressure has been normalized by $p_s$ for that case, so what is actually shown is the surface pressure anomaly.  The surface pressure is higher at the poles than at the equator, which is opposite of terrestrial-type Solar System bodies (Earth, Mars, Venus, and Titan).  As $p_s$ increases, the fractional difference between pole and equator decreases.  In the $p_s=10^4$ and $10^5$~Pa cases, the surface pressure variation is more longitudinally symmetric, with an entirely positive anomaly poleward of $\sim40-60^{\circ}$ and an entirely negative anomaly equatorward of this latitude circle.  In the $p_s=10^6$ and $10^7$~Pa cases, the surface pressure anomaly is entirely positive poleward of this latitude circle, but equatorward there are both positive and negative anomalies for a given latitude circle.  The minimum surface pressure anomaly moves poleward and westward as $p_s$ is increased; the maximum surface pressure anomaly moves mainly equatorward.

\begin{figure}
 \noindent\includegraphics[width=1.0\textwidth,angle=270]{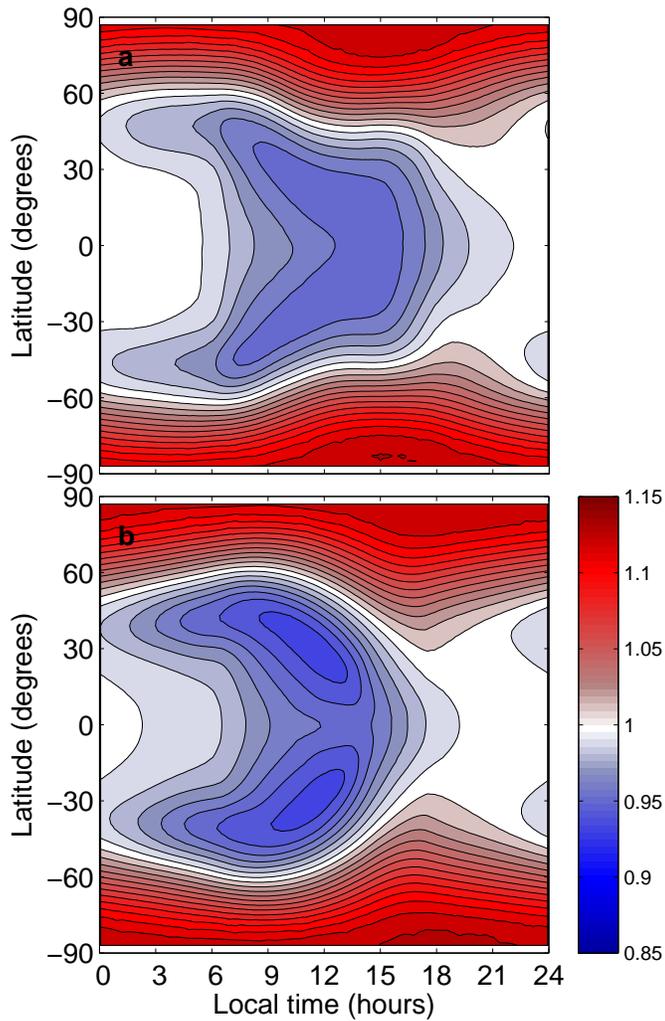}
 \caption{\label{fg:psa} GCM surface pressure variation as a function of surface albedo: (a) $A=0.1$, (b) $A=0.7$.  All cases have $p_s=10^5$~Pa and are time averaged over 400 Earth-days.  Each panel is normalized by the global mean surface pressure for that case.  Contour interval is 1\% of the global mean surface pressure.  Taken with panel (b) of Fig.~\ref{fg:psps}, these plots form a progression of surface albedos from 0.1--0.7.}
 \end{figure}

	Figure~\ref{fg:psa}, along with panel (b) of Fig.~\ref{fg:psps} (in the order Fig.~\ref{fg:psa}a, Fig.~\ref{fg:psps}b, Fig.~\ref{fg:psa}b), show the effect of increasing surface albedo on the surface pressure anomaly field.  The minimum negative anomaly deepens and moves poleward and westward, but otherwise the effect of changing surface albedo is small on the surface pressure field.
	
	Within the boundary layer (here for the case of $\sigma$ level 0.9377), as in the free atmosphere, temperature increases as $p_s$ increases.  The temperature maximum is located on the equator and at the substellar point (12 hours LT), although the contours are elongated in the eastward direction.  The temperature minimum is also located on the equator at 0 hours LT, except in the $p_s=10^7$~Pa case where it is at $\sim 40^{\circ}$ latitude.  As in the free atmosphere, temperature increases as $A$ decreases.
	
	In the boundary layer, a localized area of westerly wind exists between about 3 and 15 hours LT and 40\de~latitude.  The maximum speed in this area decreases with increasing $p_s$ and $A$.  Poleward of 40\de~latitude there is an easterly jet that is most prominent in the low $p_s$ and $A$ cases.  As both $p_s$ and $A$ increase, the jet strength decreases and the maximum speed shifts locations to the equator at 18 hours LT.  The meridional wind field is complicated, but is generally a mirror image of itself reflected across the equator.  A clear trend does not exist as $p_s$ and $A$ are increased, except that in the $p_s=10^7$ case there is an abrupt amplification of wind speeds.

\section{Discussion}\label{se:discussion}

\subsection{\citet{showman:2011} analytic shallow water model}\label{ss:sp11}

	We pay special attention to the \citet{showman:2011} shallow water analytic models, as they have already laid the analytical groundwork that will help explain our model results in simpler terms.  The \citet{showman:2011} model is a generalization of the \citet{gill:1980} and \citet{matsuno:1966} models to cases where the radiative and frictional timescales are not equal, as is the case for tidally locked, synchronously orbiting, strongly irradiated, short-period exoplanets.  This type of model leads to a solution where the day-night thermal forcing generates standing, planetary-scale Rossby and Kelvin waves, which interact with the mean flow.  The Kelvin waves span the equator, and the Rossby waves lie at the poleward boundaries of the Kelvin waves.  The combination of westerly propagating Kelvin waves; easterly propagating Rossby waves; and the pressure gradient, Coriolis, advective, and drag forces, the velocity solutions resemble eastward pointing ``chevrons'' at the equator.  These velocity tilts transport westerly momentum from high latitudes to the equator, creating the equatorial superrotation.
	
We find that the geopotential (i.e. temperature) field of their non-linear solution with radiative and drag constants of one day most closely matches our temperature pattern qualitatively (see Figs.~\ref{fg:tps16} and \ref{fg:ta16}); that is, an eastward phase shift of the temperature at the equator and a westward phase shift at high latitudes.  The temperature contours exhibit northwest-southeast tilts in the Northern hemisphere and southwest-northeast tilts in the Southern hemisphere.  \citet{showman:2011} describe this as an eastward pointing chevron pattern.  Furthermore, the absolute temperature maxima and minima are located off the equator.  This behavior can be seen in the zonally averaged potential temperature plots  (Fig.~\ref{fg:thetamean}) at $\sigma$ level 0.068 (i.e., the $\sigma$ level of Figs.~\ref{fg:tps16} and \ref{fg:ta16}), where, scanning from pole to equator, the potential temperature first increases slowly, then drops more sharply in the vicinity of the equator.   Note that a more appropriate case for comparison with our GCM would be a radiative time constant of 10 days and drag constant of one day, but \cite{showman:2011} do not show this case.
		
	The equatorial Rossby radius of deformation is given by
	\begin{equation}
	L_R=\left(\frac{\sqrt{gd}}{\beta}\right)^{1/2},
	\end{equation}
	where $g$ is the gravitational acceleration at the surface, $d$ is the shallow-water approximation depth (taken to be the depth of the model below the Rayleigh sponge layer, $10^3$~km) and $\beta=2\Omega\cos \phi /a$, where $a$ is the planetary radius.   In the \citet{showman:2011} model, the latitudinal extent of the equatorial jet is $\pm L_R$.  In our GCM, $L_R=6550$~km or $22^{\circ}$, which is smaller than the jet width of $\sim30$--$45^{\circ}$  (see Fig.~\ref{fg:umean}).  Note that the length scale $L_R$ is well resolved by our horizontal model grid. 
	
	One key difference between the \citet{showman:2011} shallow water model and our GCM \citep[and hot Jupiter models, see][and references therein]{showman:2011} is that in our GCM, the winds are westerly at all longitudes at the equator, not divergent from near the substellar point.  \citet{showman:2011} state that the shallow water (i.e., two layer) model does not capture the full multilayer structure of a 3D atmosphere.  They do however note that during the spinup phase of the \citet{showman:2009} hot Jupiter GCM, the equatorial wind field is divergent like the shallow water model, before transitioning to the fully spun up case where the zonal winds are westerly at all equatorial longitudes.  We find precisely the same behavior.  At 30 days of integration, our model shows the divergent zonal wind pattern---easterly winds from 0 to 12 hours local time and westerly winds from 12 to 24 hours local time.  At 60 days of integration, the region of easterly winds at the equator has shrunk to 6 to 15 hours local time, and by 90 days (which from Fig.~\ref{fg:kespinup} is now entering the fully spun up phase), the equatorial winds are entirely westerly.

	\subsection{Comparison with previous exoplanet GCMs}\label{ss:comparison}
	
Here we focus primarily on a comparison with super-Earth/GJ~1214b models that most closely resemble our model configuration \citep{joshi:1997,merlis:2010,menou:2012} and parameter sweep studies (Heng, Menou, and Phillipps, 2011; Heng and Vogt, 2011).  These models have been already summarized in Section~\ref{ss:gcmsummary}.
	
Our results do not compare well with the terrestrial exoplanet GCM of \citet{joshi:1997}, if we consider the $10^5$~Pa surface pressure cases (their Run~1).  Their GCM produces one overturning cell in each hemisphere, whereas our GCM produces 2.  Their temperature maximum is at the equator and the 0.8 $\sigma$ level.  Their temperature decreases monotonically towards the poles; our temperature maxima are offset from the equator.  The temperature magnitudes do not match, although this may be accounted for by the fact that we are using a much higher incoming stellar flux.  Their zonal wind pattern shows westerly flow everywhere, that increases with height and decreases from equator to pole.  Meanwhile, our GCM shows the distinct equatorial westerly jet at the equator and easterly flow at high latitudes.  Their stationary horizontal eddy momentum flux (equivalent to our Fig.~\ref{fg:etmean}) has much more structure, whereas ours are simply negative in the Northern Hemisphere and positive in the Southern Hemisphere.  We speculate that these differences are due to their assumed rotation period of one Titan day (about 16~Earth-days), compared with our rotation period of 1.58~Earth-days.  The \citet{joshi:1997} more closely resembles in a qualitative sense the slowly rotating case of \citet{merlis:2010}, which has one overturning cell in each hemisphere.
	
The quickly rotating case of \citet{merlis:2010}, with a rotation period of 1 Earth-day, $10^5$~Pa surface pressure, and surface albedo of 0.38 has the most similar configuration to our $10^5$~Pa surface pressure and 0.4 surface albedo case.  Their surface temperature pattern, while not being a quantitative match since they assumed terrestrial levels of insolation, looks like the chevron pattern of our GCM and \citet{showman:2011}, having the maximum offset from the equator and a local minimum on the equator.  However, their zonal winds at high altitudes are actually easterly at the equator (except for weak westerlies just east of substellar point) and westerly at high altitudes.  This pattern is completely opposite from our GCM and \citet{showman:2011}.  Furthermore, the \citet{merlis:2010} GCM results have three overturning cells in each hemisphere, compared with two in our GCM.  We suggest two possible causes for this discrepancy: either the slightly faster rotation rate causes the atmospheric circulation to enter a different regime (it is a general rule for Solar System planets that faster rotation rates produce more overturning cells) or there is feedback with the precipitation scheme in their model.  While they do not include the radiative effects of clouds, latent heating is present that may be adding an additional heating or cooling source to the model.  Furthermore, the winds may transport water vapor; thus the circulation and latent heating are strongly coupled.

The	configuration of the GCM of \citet{menou:2012} is slightly different than our GCM in that they do not include a surface density discontinuity; i.e., their model of GJ~1214b is a mini-Neptune class rather than a super-Earth class.  They assume an entirely \htwoo~vapor atmosphere with the lowest layer depth at $10^6$~Pa.  Their temperatures are slightly cooler than ours, and do not have an inversion in the lowest layer.  Their equatorial temperature profiles are isothermal to about the 0.05 $\sigma$ level.  These differences are almost certainly due to the different external heating/cooling schemes (i.e. RT vs. Newtonian relaxation as well as the specific dependence of opacity on pressure) between the two models.  Their zonally averaged winds are westerly everywhere except at the deepest levels.  They do have an equatorial jet, but at high latitudes the winds are still westerly, compared with easterly in our GCM.  This is possibly a secondary consequence of the different external heating/cooling schemes, but may also have to do with the drag at the bottom levels.  They parameterize magnetic drag while we parameterize frictional drag, but in any case their drag time scale is 20 and 40 days in the lowest two model levels, while ours decreases from 1 day at the lowest level to 0 at the 0.7 $\sigma$ level.

Heng, Menou, and Phillipps (2011) use the \citet{held:1994} benchmark as their equilibrium temperature specification, modified to include the  geometric effects of tidal locking and synchronous rotation (Eq.~\ref{eq:hs}), to simulate a hypothetically tidally locked and synchronously orbiting Earth (surface pressure $10^5$~Pa).  They use a rotation rate equal to one Earth-year, as in the slowly rotating case of \citet{merlis:2010}.  Not unexpectedly, their results  do not match ours, having a surface temperature maximum located at the substellar point and zonal winds that are divergent from the substellar point.

Heng and Vogt (2011) perform a parameter sweep over many different GCM parameters; the one in common with our parameter sweep is the global mean surface pressure.  Heng and Vogt (2011) use Newtonian relaxation to the equilibrium temperature specification from \citet{held:1994} (modified to include the geometric effects of tidal locking and sychronous orbit, Eq.~\ref{eq:hs}).  Their rotation rate is 37 Earth-days, compared to 1.58 Earth-days in our simulations.  Their results for zonally averaged winds are very reminiscent of \citet{joshi:1997} Run~1; that is, westerly everywhere, increasing with height, and decreasing from equator to pole.  Heng and Vogt (2011) conclude that there is no qualitative difference in the zonal mean zonal wind profiles for the cases they tested ($10^4$, $10^5$, $3\times10^5$, and $10^6$~Pa).  We speculate that this is because they are in the slowly rotating regime, while we are in the rapidly rotating regime, and it is this point that makes our parameter sweep novel.

We finally note that our GCM shares the entirely westerly equatorial jet shown by certain other fully spun up hot Jupiter and super Earth GCMs (e.g. Showman and Guillot, 2002; Cooper and Showman, 2005; Showman et al., 2008, 2009; Heng, Menou, and Phillipps, 2011; Rauscher and Menou, 2012).  This seems to be a robust feature of synchronously rotating exoplanets due to their subsequent intense day-night heating gradient~\citep{showman:2011}, with the caveat that they must be quickly (with periods of order 1 Earth-day) rotating.  It may also be due to the fact that our model domain extends much higher than most Earth or super-Earth models.

		\subsection{Transmission spectra}\label{ss:spectra}
	
	We compute model atmospheric spectra from our GCM output to determine if the atmospheres we have simulated here could be distinguished by observations.  Specifically, we compute transmission spectra using a line-by-line radiative transfer code adapted from \citet{madhusudhan:2009}. Given the pressure--temperature ($P$--$T$) profiles at the limb of the atmosphere and the assumed \htwoo~composition, our spectral model computes a transmission spectrum of planet under the assumption of local thermodynamic equilibrium and hydrostatic equilibrium. The $P$--$T$ profile in our spectral model comprises of 100 layers uniformly separated in pressure between $p_s$ and 1~Pa (i.e, $10^{-5}$ bar), interpolated between the pressure layers of the $P$--$T$ profiles we adopt from the GCM output. In cases where our GCM output does not extend to low enough pressures, the atmosphere is assumed isothermal with the temperature set equal to the temperature of the uppermost GCM level below the Rayleigh friction layer. This is roughly valid for two reasons. First, at low pressures the temperature profile is expected to asymptote to an isotherm  (i.e., skin temperature). Second, the actual temperature may be off by $\sim$100~K or less from the correct profile, but transmission spectra for exoplanets are presently insensitive to such temperature differences. 

\begin{figure}
 \noindent\includegraphics[width=1.0\textwidth]{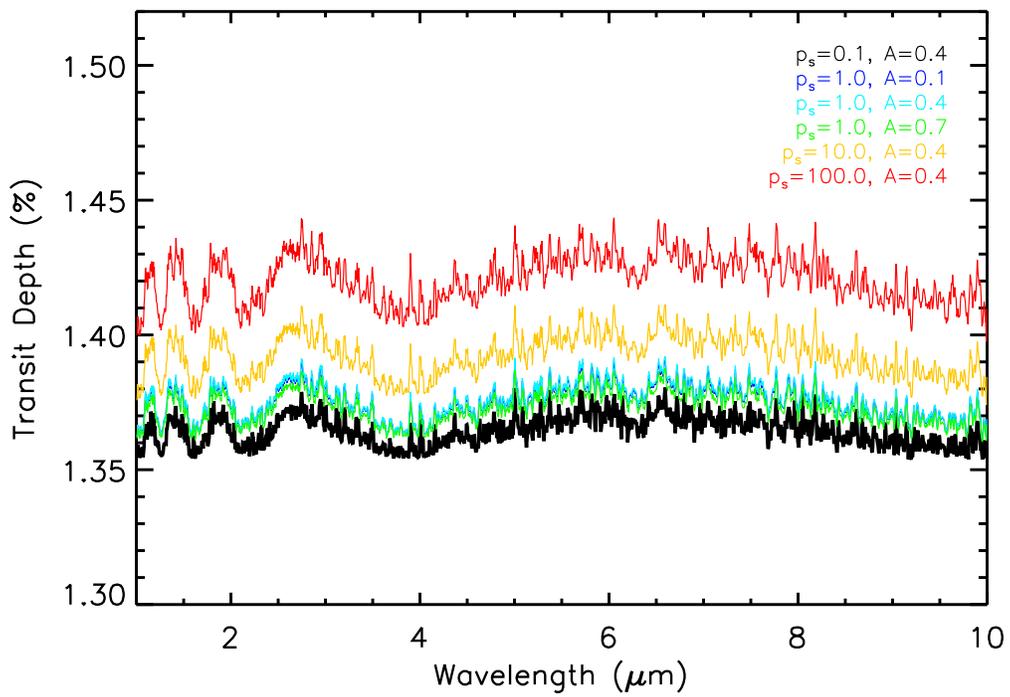}
 \caption{\label{fg:spectra} Model \htwoo~transmission spectra calculated from GCM pressure--temperature output. The units for $p_s$ are in bars (1~bar = $10^5$~Pa).}
 \end{figure}

Figure~\ref{fg:spectra} shows our model transmission spectra for the different cases we consider in this work. Infrared spectra are shown for different global mean surface pressures and surface albedos. As shown in Fig.~\ref{fg:spectra}, the infrared transit depth increases with increasing $p_s$ over the entire wavelength range between 1 and 10 $\mu$m. This is simply because with increasing surface pressure the infrared photosphere (the radius below which effectively all incident starlight is blocked from the observer) of the planet is reached at a higher altitude, causing a higher effective radius of the planet.  Current observations would not be able to distinguish variations in surface albedo of our hypothetical planet.
Observations with existing instruments may however be able to distinguish between atmospheres varying in surface pressure
by over a decade (e.g.  between 0.1 and 10 bar, 1 and 100 bar, and 10 and 100 bar). As shown in Fig.~\ref{fg:spectra}, such differences
in surface pressure translate to $\geq$200 ppm differences in transit depths across the near to mid infrared
wavelengths, and hence are detectable with current instruments which routinely obtain precisions of 100 ppm and less \citep{bean:2011}.

The differences between the spectra from models that differ by more than two decades in $p_s$ are observable with current instruments (i.e. 0.1 and 10~bar, 1 and 100 bar, etc.). For higher $p_s$, even models with $p_s = 10$ bar and $p_s = 100$~bar are distinguishable. The variation in the infrared transit depth as a function of albedo is smaller than the variation with pressure.  

\section{Conclusions}
We have simulated, using a GCM with a Newtonian relaxation scheme, a hypothetical, tidally locked, synchronously orbiting super-Earth under the following assumptions: (1) the size, mass, and orbital characteristics of GJ~1214b, (2) a greenhouse-gas dominated atmosphere, (3) the gas properties of \htwoo, and (4) a surface.  We have performed a parameter sweep in global mean surface pressure ($10^4$, $10^5$, $10^6$, and $10^7$~Pa) and global mean surface albedo (0.1, 0.4, and 0.7).  Our study differs from the parameter sweep of Heng and Vogt (2011) in that we are in the rapidly rotating planet regime (diurnal period of 37 Earth-days vs. 1.58 Earth-days), which has been explicitly shown to be markedly different from the slowly rotating regime examined by \citet{merlis:2010}.  The other difference in our study from Heng and Vogt (2011) is that our relaxation scheme relaxes to a true analytical (albeit gray) solution of the RT equation, instead of the \citet{held:1994} specification.  While the \citet{held:1994} equilibrium temperature is the golden standard for benchmark studies of the terrestrial atmosphere, it is an intentionally prescribed function that reproduces Earth's atmosphere, and therefore may not be general enough for other planets.

An equatorial, westerly, superrotating jet is a robust feature in our GCM.  At high latitudes, the flow is easterly.  One key point is that the equatorial jet is westerly at all longitudes.  This may be reconciled with the \citet{showman:2011} analytical models, which is a generalization of the \citet{gill:1980} and \citet{matsuno:1966} shallow water models, by noting that during spinup phase, full 3-D GCMs~\citep[e.g.][]{showman:2009} do show divergent zonal flow at the equator (interpreted as easterly and westerly propagating Rossby and Kelvin waves, respectively).  However, when the steady state is reached, the equatorial flow becomes entirely westerly and the analogy with the shallow water model breaks down, which \citet{showman:2011} suggest is due to the multi-layered configuration of 3-D GCMs.  Our GCM results reproduce this behavior precisely.

Unlike Heng and Vogt (2011), our zonal winds do show a change with global mean surface pressure.  As global mean surface pressure increases, the speed of the equatorial jet decreases between 9 and 15 hours LT.  The latitudinal extent of the equatorial jet increases on the nightside.  For the two greatest initial surface pressure cases ($10^6$ and $10^7$~Pa), an increasingly westerly component of flow develops at middle to high latitudes between 11 and 18 hours LT.  The nightside, easterly flow in the midlatitudes also increases in speed as global mean surface pressure increases.  Furthermore, the zonal wind speed in the equatorial and midlatitude jets decreases with increasing surface albedo.  Also, the latitudinal width of the equatorial jet decreases as surface albedo increases.

This study, i.e., a parameter sweep in the rapidly rotating super-Earth regime, could be repeated after several model improvements such as replacing the Newtonian relaxation scheme with directly calculated heating and cooling rates, a state-of-the-art (e.g. double-grey, dual-band, or correlated-k) radiation scheme, surface/subsurface model, boundary layer scheme, and condensation (i.e. surface-atmosphere mass exchange) cycle.  All of these features have already been included in many exoplanet models; it is merely a matter of integrating them into our present model.


\section*{Acknowledgments}
We thank two anonymous referees for detailed comments leading to substantial revision of this work.


\appendix

\renewcommand{\theequation}{A-\arabic{equation}}
  \setcounter{equation}{0}  
  \section*{Appendix: The validity of the hydrostatic approximation}
The vertical momentum equation (neglecting friction) is
\begin{equation}\label{eq:vm}
\frac{Dw}{Dt}-2\Omega u cos\phi - \frac{\left(u^2+v^2\right)}{a} = - \frac{1}{\rho}\frac{\partial p}{\partial z} - g,
\end{equation}
where $\rho$ is the atmospheric density and $z$ is altitude.  Using scale analysis techniques~\citep[e.g.][]{holton:1992}, Eq.~\ref{eq:vm} may be rewritten by its scaling terms as
\begin{equation}\label{eq:vmscale}
\frac{UW}{L_o} - f_o U - \frac{U^2}{a_o} = - \frac{P_o}{\rho_o H} -G,
\end{equation}
where $U$ is a characteristic velocity scale (taken to be 100~\ms), $W$ is a characteristic vertical velocity scale in height coordinates (taken to be 0.1~\ms), $L_o$ is a characteristic horizontal length scale (taken to be $10^7$~m), $f_o$ is the average Coriolis parameter (taken to be $10^{-4}$~s$^{-1}$), $a_o$ is the order-of-magnitude planetary radius (taken to be $10^7$~m), $P_o$ is the surface pressure (here we use the nominal case of $10^5$~Pa), $\rho_o$ is the representative surface density (here taken to be for the $10^5$~Pa case, 1~kg~m$^3$), $H$ is a representative vertical depth (here taken to be the scale height for the $10^5$~Pa case, 10~km), and $G$ is an order of magnitude for the surface gravitational acceleration (taken to be 10~m~s$^{-2}$).  Note that, following the method of scale analysis, we are concerned with the order-of-magnitude values of the terms in Eq.~\ref{eq:vmscale}, not their precise values.  Then, from left to right, the values of the terms in \ref{eq:vmscale} are: $10^{-6}$, $10^{-2}$, $10^{-3}$, 10, and 10.  This indicates that the two terms on the right hand side of \ref{eq:vm} are by far the largest terms, i.e., the hydrostatic balance approximation is valid.  Note that the first term on the right hand side depends on surface pressure, which we vary in our parameter sweep.  However, as surface pressure increases, so will $\rho_o$ and $H$ due to the increased greenhouse effect, and this term will remain the same order of magnitude.  The hydrostatic balance approximation is then still valid.


\nocite{heng:2011}
\nocite{heng:2011a}
\nocite{heng:2011b}


\label{lastpage}

\clearpage	

\begin{table}[h]
\centering
\caption{Selected GCM parameters}

\begin{tabular}{lr}\label{tb:gcmall}

\\
\hline \hline
Parameter & Value \\
\hline
Surface gravitational acceleration, $g$ & 8.93  m~s$^{-1}$ \\
Orbital eccentricity & 0  \\
Orbital semimajor axis & 2.1426$\times 10^{9}$  m  \\
Stellar mass & 3.1243 $\times 10^{29}$  kg \\
Rotation rate ($\Omega$) & 4.602$\times 10^{-5}$  s$^{-1}$  \\
Obliquity & 0  $^{\circ}$  \\
Solar constant & 21519  W~$m^{-2}$  \\
Surface emissivity, $\epsilon$ & 1.0  \\
Radius & 17059  km \\
Friction time scale ($1/k_f$) & 1 Earth-day \\
Specific heat at constant pressure  & 1850 J~kg$^{-1}$~K$^{-1}$ \\
Specific latent heat of vaporization  & 22.6 $10^5$ J~kg$^{-1}$\\
Vaporization reference pressure ($T_1$) & 1.01325 $10^5$ Pa\\
Vaporization reference temperature ($p_1$) & 373 K\\
Specific gas constant & 461 J~kg$^{-1}$~K$^{-1}$ \\
Ratio of specific gas constant to specific heat & 1/4   \\
Long wave column optical depth at $10^5$~Pa ($\tau_{oo}$)  & 1.0  \\
Radiative relaxation time scale ($1/k_T$) & 12.6 Earth-days  \\

\hline

\end{tabular}
\end{table}

%
%
%
%
%

\clearpage


%

 \clearpage
 
\end{document}